\documentclass[12pt]{article}
\usepackage{epsf}
\usepackage{epsfig}
\usepackage{axodraw}

\def\beq{\begin{eqnarray}}
\def\eeq{\end{eqnarray}}

\begin{document}

\begin{titlepage}
\begin{flushright}
 UAB-FT-536\\
 PITHA 02/20\\
 hep-ph/0212158\\
 December 2002
\end{flushright}
\vskip 2.0cm

\begin{center}
{\Large\bf Forward-backward and isospin asymmetry 
\\[0.3em] for $B \to K^\ast \ell^+ \ell^-$ decay
\\[0.4em] in the standard model and in supersymmetry
}
%---------------------------------------------------------
\vskip 2.2cm
{\sc Th.~Feldmann}
\vskip .5cm
{\it Institut f\"ur Theoretische Physik~E, RWTH Aachen,
52056 Aachen, Germany}
\vskip 0.7cm
and
\vskip 0.7cm
\hspace*{0.3cm}{\sc J.~Matias}
\vskip .5cm
{
%\it Theory Division, CERN, CH-1211 Geneve 23, Switzerland \\
%and \\
\it IFAE, Universitat Aut\`onoma de Barcelona,  08193 Bellaterra,
 Barcelona, Spain}

\vskip 1.5cm

\end{center}

\begin{abstract}
\noindent
We discuss two dedicated observables in exclusive
$B \to K^\ast \ell^+ \ell^-$ decay that can be used to study
effects of physics beyond the standard model, namely
the forward-backward asymmetry in the lepton spectrum and
the isospin-asymmetry between decays of charged and neutral
$B$~mesons. We consider the region of large recoil-energy (i.e.\
small invariant mass of the lepton pair),
and employ the QCD factorization approach to exclusive $B$~meson decays.
Sub-leading effects in the heavy quark mass expansion have been taken
into account for the calculation of the isospin-asymmetry. 
We give predictions for decay asymmetries in the
standard model, and its supersymmetric extension with
minimal flavor violation, using parameter values allowed
by current experimental constraints on $B \to X_s\gamma$ decay.
\end{abstract}

\vfill

%\centerline{\it (to be submitted to JHEP)}

%\vfill

\end{titlepage}

%%%%%%%%%%%%%%%%%%%%%%%%%%%%%%%%%%%%%%%%%%%%%%%%%%%%%%%%%%%%%%%%%%%%%%%

\section{Introduction}

$B$~physics has become one of the most active fields in the search for new
interactions associated to physics beyond the standard model (SM).
In particular, the flavour sector of the new theory should have
distinctive effects on exclusive and inclusive rare $B$~decays 
which are suppressed in the SM.
At present, one of the most important constraints on
new physics from rare $B$~decays arises from the inclusive decay $B 
\rightarrow X_s \gamma$.

The inclusive decay processes $B \rightarrow X_s \gamma$ and also $B
\rightarrow X_s l^+l^-$ have been studied in great detail in the
literature in the SM \cite{smbsg}-\cite{smbsll} and MSSM 
\cite{bbmr}-\cite{susybsll3}. Concerning $B \to X_s \gamma$ we have at our
disposal accurate determinations both from experimental~\cite{expbsgamma}
and theoretical point of view~\cite{smbsg}-\cite{Buchalla:1998ky}.
Within the present uncertainties $B \to X_s \gamma$ is consistent with
the SM prediction, which translates into 
a quite narrow allowed region for the Wilson coefficient $C_7^{\rm eff}$ in
the weak effective Hamiltonian  for
$b \to s\gamma$ (also $b \to s \gamma \gamma$~\cite{Bertolini:1998hp}) and 
$b\to  s\ell^+\ell^-$ transitions.
The inclusive $B \to X_s l^+ l^-$ decay has not been measured yet;
for the theoretical literature we refer the reader to
\cite{smbsll},\cite{Cho:1996we}-\cite{susybsll3}.

In this paper we will focus on the exclusive decay mode
$B \rightarrow K^\ast l^+ l^-$.
In general, exclusive decays are much easier accessible in experiment
(first experimental evidence for these decay mode has been presented in
\cite{Walsh:2002ux,Aubert:2002pj}), 
but their theoretical treatment is usually 
more difficult because of non-perturbative hadronic binding effects.
However, in the limit where the invariant mass
of the lepton pair is small, the process admits a systematic theoretical
description using QCD factorization in the heavy quark limit
\cite{BBNS}.
As stressed in \cite{Beneke:2001at,Burdman:1998mk,Ali:1999mm,AFB},
a distinct observable in this decay is
the forward-backward (FB) asymmetry in the lepton spectrum: In particular
the FB asymmetry-zero
can be calculated with rather high accuracy since
most of the hadronic uncertainties drop out.
This leads to an almost model-independent relation
between the Wilson coefficients $C_7^{\rm eff}$ and $C_9$.

We further stress that the sensitivity to new physics in exclusive modes 
may be different from that of their inclusive counterparts 
if the spectator quark is involved in the short-distance dynamics.
This is the case for the isospin asymmetry between $B \to
K^*\ell^+\ell^-$ decays with charged or neutral $B$ mesons.
Kagan and Neubert \cite{Kagan:2001zk} have shown 
that annihilation topologies provide
the main source for the isospin asymmetry in  $B \to K^\ast\gamma$ decays,
and concluded that the sign and
magnitude of the isospin asymmetry is predominantly
determined by the Wilson coefficients
$C_5$ and $C_6$ of four-quark penguin operators and their
relative sign with $C_7^{\rm eff}$. 
The SM prediction is consistent with experiment, 
within still large uncertainties.
In this article we will extent the analysis of  \cite{Kagan:2001zk}
to the decay $B\to K^\ast\ell^+\ell^-$  and calculate the
isospin-asymmetry for small invariant lepton-pair masses 
($q^2$ below $7$~GeV$^2$).

The paper is organized in the following way.
In the next section we will shortly review 
the FB asymmetry for the decay 
$B \rightarrow  K^\ast l^+l^-$ within the QCD factorization approach.
Then we turn to the calculation of the isospin asymmetry in 
$B \rightarrow  K^\ast l^+l^-$, and give a numerical estimate within the SM.
In section~3 we explore new physics contributions
to both observables, the FB and the isospin-asymmetry,
in the MSSM with minimal flavor violation (MFV). We apply the
constraints from $B \to X_s\gamma$ using the NLO calculation in
\cite{Ciuchini:1998xy,Degrassi:2000qf}.
Section~4 is devoted to a brief discussion of possible effects
beyond MFV, before we conclude with a summary in section~5.

\section{$B \rightarrow K^\ast \ell^+ \ell^-$ decay asymmetries
  in the SM}

\label{sec2}

The decay $B \to K^\ast \ell^+ \ell^-$
is induced by a set of operators ${\cal O}_i$ appearing in the
weak effective Hamiltonian for $b\to s$ transitions
\cite{Buchalla:1996vs} 
\beq \label{heff}
  H_{\rm eff} &=& \frac{G_F}{\sqrt2} \,
  \left\{ \sum\limits_{i=1}^2
   (\lambda_u \,  C_i \, O_i^u + \lambda_c \, C_i \, O_i^c)
   - \lambda_t \, \sum\limits_{i=3}^{10}  C_i \, O_i \right\} \, .
\eeq
Here, the combinations of CKM matrix elements are abbreviated as
$\lambda_q = V_{qs}^* \, V_{qb}$, and obey the unitarity
relation $\lambda_u + \lambda_c = - \lambda_t$.
The Wilson coefficients $C_i$ contain the short-distance 
physics related to integrating out heavy particles (like 
weak gauge bosons and the top quark) and hard gluons.
In extensions of the SM new (heavy) particles 
modify the SM values of Wilson coefficients and/or lead to
new operators in addition to (\ref{heff}). In this paper we will restrict
ourselves to the case of SM operators (\ref{heff}) only.
(We will use the same conventions as in \cite{Beneke:2001at}; in particular,
the gauge vertices read $+i e e_q$ and
$+ig_s T^A$,  and the SM values for $C_7^{\rm eff}$ and
$C_8^{\rm eff}$ are negative.)

%%%%%%%%%%%%%%%%%%%%%%%%%%%%%%%%%%%%%%%%%%%%%%%%%%%%%%%%%%%%%%%%%%%%%%%

A systematic calculation of QCD effects in exclusive $B$~decays
requires the factorization of long- and short-distance physics
distinguished by the heavy quark mass $m_b$
in the hadronic matrix elements of $H_{\rm eff}$.
The theoretical framework has been introduced for
non-leptonic $B$~decays by Beneke/Buchalla/Neubert/Sachrajda (BBNS)
\cite{BBNS}, and has been
extended to heavy-to-light form factors and radiative $B$~decays
in 
\cite{Beneke:2001at}, \cite{Kagan:2001zk}-\cite{Descotes-Genon:2002mw}.
%,Beneke:2000wa,Bosch:2001gv,Ali:2001ez,Bosch:2002bv,
In the BBNS approach, starting from the heavy quark limit,
$m_b \to \infty$, radiative corrections to 
``naive'' factorization of hadronic amplitudes arising
from hard gluon exchange can be calculated in terms
of perturbative hard-scattering kernels and universal 
(non-perturbative) hadronic quantities (decay constants,
form factors, and light-cone distribution amplitudes).

In order to constrain new-physics
scenarios in exclusive decays with reasonable accuracy,
one has to find observables that are sensitive to the short-distance
physics and rather independent to
hadronic input parameters like form factor values, moments
of light-cone wave functions etc.
In this paper we will focus on two asymmetries in the lepton spectrum,
namely the forward-backward asymmetry
\begin{eqnarray}
 \frac{dA_{\rm FB}}{dq^2} &\equiv &
 \frac{1}{d\Gamma/dq^2} \left(\,
\int_0^1 d(\cos\theta) \,\frac{d^2\Gamma[B \to K^\ast \ell^+ \ell^-]}{dq^2 d\cos\theta} -
\int_{-1}^0 d(\cos\theta) \,\frac{d^2\Gamma[B \to K^\ast \ell^+ \ell^-]}{dq^2 d\cos\theta}
\right) \, ,
\cr &&
\label{dAFB}
\end{eqnarray}
and the ($CP$ averaged) isospin asymmetry
\begin{eqnarray}
\frac{dA_I}{dq^2} &\equiv & 
\frac{ d\Gamma[B^0\to K^{\ast0}\ell^+ \ell^-]/dq^2 -
d\Gamma[B^\pm\to  K^{\ast\pm}\ell^+ \ell^-]/dq^2}
{ d\Gamma[ B^0\to K^{\ast0}\ell^+ \ell^-]/dq^2 +
d\Gamma[B^\pm \to  K^{\ast\pm}\ell^+ \ell^-]/dq^2} \ .
\label{dAI}
\end{eqnarray}
Because of the sensitivity to relative signs and orders of magnitude
between different Wilson coefficients,
these asymmetries are particularly useful to constrain 
new physics scenarios. (\cite{Ali:2002qc} entertains another
possibility to study physics beyond the SM in this $B$ decay mode,
by comparing helicity amplitudes
for $B \to K^\ast \ell^+ \ell^-$ and $B \to \rho \ell^+ \nu$.)  

\subsection{Forward-backward asymmetry}

We will follow the results and notation of
~\cite{Beneke:2001at} where the QCD corrections to the decay
$B \to K^\ast \ell^+\ell^-$ have been calculated to leading power in
the inverse heavy quark mass or light-quark energy, respectively, and
to NLO in the strong coupling constant. The approach is restricted
to the situation where the energy of $K^\ast$ is large, i.e.\
the invariant lepton-pair mass $q^2$ is small. For practical
purposes this means that $q^2$ is below charm threshold, 
$q^2 < 4 m_c^2 \approx 7$~GeV$^2$.
The decay amplitude for $B \to K^\ast\ell^+\ell^-$ is
conveniently expressed in terms of the
quantities
${\cal C}_9^\perp(q^2)$ and ${\cal C}_9^\parallel(q^2)$,
which are generalizations of the effective Wilson coefficient
$C_9^{\rm eff}$ used in the ``naive'' factorization approach.
The lowest-order (naively factorizing) expressions  
read
\beq
  {\cal C}_9^{(0)\,\perp}(q^2) &=& C_9 + Y(q^2) + \frac{2 m_b M_B}{q^2}
  \, C_7^{\rm eff} \ ,
\cr
  {\cal C}_9^{(0)\,\parallel}(q^2) &=& C_9 + Y(q^2) + \frac{2 m_b }{M_B}
  \, C_7^{\rm eff} \ .
\label{C9fac}
\eeq
(The function $Y(s)$ can be found, for instance,
 in \cite{Buchalla:1996vs,Beneke:2001at}).
The differential decay rate then reads
\cite{Beneke:2001at}
\begin{eqnarray}
\frac{d^2\Gamma}{dq^2 d\!\cos\theta}
&=& \frac{G_F^2 |V_{ts}^*V_{tb}|^2}
{128\pi^3}\,M_B^3\,\lambda(q^2,m_{K^\ast}^2)^3
\left(\frac{\alpha_{\rm em}}{4\pi}\right)^{\!2}\nonumber\\
&&
\Bigg[(1+\cos^2\theta)\,\frac{2 q^2}{M_B^2}\,\xi_\perp(q^2)^2
\left(\,|{\cal C}_9^\perp(q^2)|^2+C_{10}^2\right)
\nonumber\\
&&
+\,(1-\cos^2\theta)\,
\left(\frac{E\,\xi_\parallel(q^2)}{m_{K^\ast}}\right)^{\!2}\,
\left(\,|{\cal C}_9^\parallel(q^2)|^2+C_{10}^2\,\Delta_\parallel(q^2)^2\right)
\nonumber\\
&&-\,\cos\theta\,\frac{8 q^2}{M_B^2}\,\xi_\perp(q^2)^2\,
\mbox{Re}({\cal C}_9^\perp(q^2)) \,C_{10}
\Bigg] \ .
\label{dGamma}
\end{eqnarray}
Here
\begin{equation}
\lambda(q^2,m_{K^\ast}^2) = \Bigg[\left(1-\frac{q^2}{M_B^2}\right)^2-
    \frac{2 m_{K^\ast}^2}{M_B^2}
\left(1+\frac{q^2}{M_B^2}\right)+\frac{m_{K^\ast}^4}{M_B^4}\Bigg]^{\!1/2}
\end{equation}
is the usual kinematic factor,
$m_{b,\rm PS}$ is the potential-subtracted heavy
quark mass~\cite{BEneke:1998rk},
$E \simeq (M_B^2-q^2)/2M_B$ is the $K^\ast$ energy,
and $\theta$ refers to the angle between the positively charged lepton and the
$B$~meson in the center-of-mass frame of the lepton pair.
Furthermore $\xi_\perp(q^2)$ and
$\xi_\parallel(q^2)$ are the universal soft heavy-to-light
form factors, arising in the heavy-quark and large-recoil-energy limit
\cite{Charles:1998dr,Beneke:2000wa} (for a critical examination
whether the form factors themselves can be calculated perturbatively,
see~\cite{Descotes-Genon:2001hm} and references therein.) The function
$\Delta_\parallel(q^2)$ is given in~\cite{Beneke:2001at}.

\begin{figure}[t]
\begin{center}
\epsfclipon
\psfig{bb=135 520 375 675,file=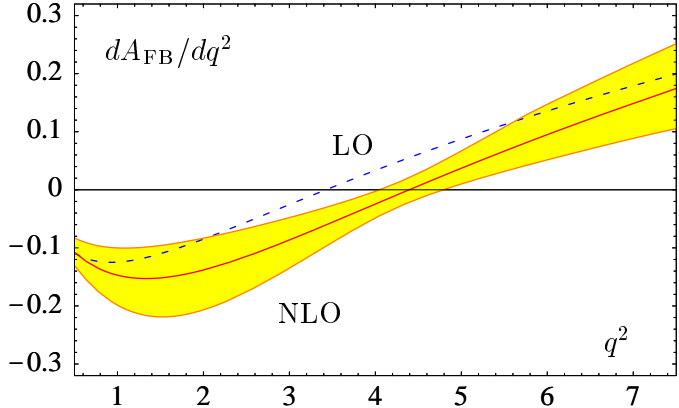, width=0.6\textwidth}
\end{center}
\caption{\label{fig:dafbnorm}
The SM forward-backward asymmetry
$d A_{\rm FB}/dq^2$ for the decay $B^-\to K^{\ast-}\ell^+\ell^-$
as a function of the invariant lepton-pair mass $q^2$
at next-to-leading order (solid center line) and leading order
(dashed). The band reflects all theoretical uncertainties from
input parameters and scale dependence combined.
Figure taken from \cite{Beneke:2001at}. }
\end{figure}

{}From (\ref{dGamma}) it is obvious that
the FB asymmetry $dA_{\rm FB}/dq^2$ in
(\ref{dAFB}) is proportional to 
$\mbox{Re}\left({\cal C}_{9}^\perp(q^2)\right)$,
and therefore it vanishes if
$\mbox{Re}\left({\cal C}_{9}^\perp(q_0^2)\right)=0$.
At leading order this gives a relation between the short-distance
Wilson coefficients $C_9$ and $C_7^{\rm eff}$ which is completely
independent of hadronic uncertainties
\cite{Burdman:1998mk,Ali:1999mm}
\begin{eqnarray}
\mbox{LO:} \qquad C_9 +  \mbox{Re}(Y(q_0^2))
 &=&  - \frac{2 M_B m_b}{q_0^2} \, C_7^{\rm eff} \ .
\end{eqnarray}
Therefore the measurement of the asymmetry zero provides a crucial
test of the SM and new physics.
(Phenomenological consequences for the FB asymmetry
from physics beyond the SM have already been discussed
in~\cite{Ali:1999mm} using ``naive'' factorization.)
In Fig.~\ref{fig:dafbnorm} we plot the results for the
FB asymmetry obtained in~\cite{Beneke:2001at}
for SM Wilson coefficients.
A conservative estimate for the location of the asymmetry zero
in the SM
(taking into account the possibility of additional model-dependence
from the form factors arising at higher order in the $\Lambda_h/m_B$
expansion) was obtained as
\begin{eqnarray}
\label{q0result2}
  q_0^2 &=& (4.2\pm 0.6)\,{\rm GeV}^2 \ .
\end{eqnarray}

%%%%%%%%%%%%%%%%%%%%%%%%%%%%%%%%%%%%%%%%%%%%%%%%%%%%%%%%%%%%%%%%%%%

%%%%%%%%%%%%%%%%%%%%%%%%%%%%%%%%%%%%%%%%%%%%%%%%%%%%%%%%%%%%%%%%%%%

\subsection{Isospin asymmetry}

 \begin{figure}[bhpt]

\begin{center}
   (a) \hskip3em
       \parbox[c]{0.8\textwidth}{     
       \begin{picture}(150,90)(0,0)

       \DashLine(0,50)(25,50){6}

       \ArrowLine(25,70)(65,50)
       \ArrowLine(65,50)(45,40)
       \ArrowLine(45,40)(25,30)
       \ArrowLine(105,30)(65,50)
       \ArrowLine(65,50)(105,70)

       \DashLine(105,50)(130,50){6}

       \GOval(105,50)(20,10)(0){0.99}
       \GOval(25,50)(20,10)(0){0.99}

       \Photon(45,40)(80,20){3}{7}
       \Vertex(45,40){2}

       \GOval(65,50)(5,5)(0){0.5}

       \Text(-10,50)[]{\small $B$}
       \Text(140,50)[]{\small $K^\ast$}

     \end{picture}
     \hskip3em
     \begin{picture}(150,75)(0,0)

       \DashLine(0,50)(25,50){6}

       \ArrowLine(25,70)(65,50)
       \ArrowLine(65,50)(25,30)
       \ArrowLine(85,40)(65,50)
       \ArrowLine(105,30)(85,40)
       \ArrowLine(65,50)(105,70)

       \DashLine(105,50)(130,50){6}

       \GOval(105,50)(20,10)(0){0.99}
       \GOval(25,50)(20,10)(0){0.99}

       \Photon(85,40)(50,20){3}{7}
       \Vertex(85,40){2}

       \GOval(65,50)(5,5)(0){0.5}

       \Text(-10,50)[]{\small $B$}
       \Text(140,50)[]{\small $K^\ast$}

     \end{picture}
}

(b) \hskip3em
  \parbox[c]{0.8\textwidth}{
 \begin{picture}(150,75)(0,0)

       \DashLine(0,50)(25,50){6}

       \ArrowLine(25,70)(65,70)
       \ArrowLine(65,30)(45,30)
       \ArrowLine(45,30)(25,30)
       \ArrowLine(105,30)(65,30)
       \ArrowLine(65,70)(105,70)

       \DashLine(105,50)(130,50){6}

       \GOval(105,50)(20,10)(0){0.99}
       \GOval(25,50)(20,10)(0){0.99}

       \Photon(45,30)(80,10){3}{7}
       \Vertex(45,30){2}

       \Gluon(65,70)(65,30){3}{7}
       \Vertex(65,30){2}
       \GOval(65,70)(5,5)(0){0.5}

       \Text(-10,50)[]{\small $B$}
       \Text(140,50)[]{\small $K^\ast$}

     \end{picture}
     \hskip3em
     \begin{picture}(150,75)(0,0)

       \DashLine(0,50)(25,50){6}

       \ArrowLine(25,70)(65,70)
       \ArrowLine(65,30)(25,30)
       \ArrowLine(85,30)(65,30)
       \ArrowLine(105,30)(85,30)
       \ArrowLine(65,70)(105,70)

       \DashLine(105,50)(130,50){6}

       \GOval(105,50)(20,10)(0){0.99}
       \GOval(25,50)(20,10)(0){0.99}

       \Photon(85,30)(50,10){3}{7}
       \Vertex(85,30){2}

       \Gluon(65,70)(65,30){3}{7}
       \Vertex(65,30){2}
       \GOval(65,70)(5,5)(0){0.5}

       \Text(-10,50)[]{\small $B$}
       \Text(140,50)[]{\small $K^\ast$}

     \end{picture}
}

(c) \hskip3em
\parbox[c]{0.8\textwidth}{
     \begin{picture}(150,75)(0,0)

       \DashLine(0,50)(25,50){6}

       \ArrowLine(25,70)(65,70)
       \ArrowLine(65,30)(45,30)
       \ArrowLine(45,30)(25,30)
       \ArrowLine(105,30)(65,30)
       \ArrowLine(65,70)(105,70)

       \DashLine(105,50)(130,50){6}

       \GOval(105,50)(20,10)(0){0.99}
       \GOval(25,50)(20,10)(0){0.99}

       \Photon(45,30)(80,10){3}{7}
       \Vertex(45,30){2}

       \ArrowArc(65,58)(12,90,270)
       \ArrowArc(65,58)(12,270,450)

       \Gluon(65,46)(65,30){3}{3}
       \Vertex(65,30){2}
       \Vertex(65,46){2}
       \GOval(65,70)(5,5)(0){0.5}

       \Text(-10,50)[]{\small $B$}
       \Text(140,50)[]{\small $K^\ast$}

     \end{picture}
     \hskip3em
     \begin{picture}(150,75)(0,0)

       \DashLine(0,50)(25,50){6}

       \ArrowLine(25,70)(65,70)
       \ArrowLine(65,30)(25,30)
       \ArrowLine(85,30)(65,30)
       \ArrowLine(105,30)(85,30)
       \ArrowLine(65,70)(105,70)

       \DashLine(105,50)(130,50){6}

       \GOval(105,50)(20,10)(0){0.99}
       \GOval(25,50)(20,10)(0){0.99}

       \Photon(85,30)(50,10){3}{7}
       \Vertex(85,30){2}

       \ArrowArc(65,58)(12,90,270)
       \ArrowArc(65,58)(12,270,450)

       \Gluon(65,46)(65,30){3}{3}
       \Vertex(65,30){2}
       \Vertex(65,46){2}
       \GOval(65,70)(5,5)(0){0.5}

       \Text(-10,50)[]{\small $B$}
       \Text(140,50)[]{\small $K^\ast$}

     \end{picture}
}   \end{center}
\caption{Contributions to 
 the isospin asymmetry $dA_I/dq^2$:
(a) Annihilation topologies with
operators ${\cal O}_{1-6}$,
(b) Hard spectator interaction involving the gluonic penguin
    operator ${\cal O}_8$, (c) Hard spectator interaction 
   involving the operators  ${\cal O}_{1-6}$.}
\label{dI_fig}
\end{figure}
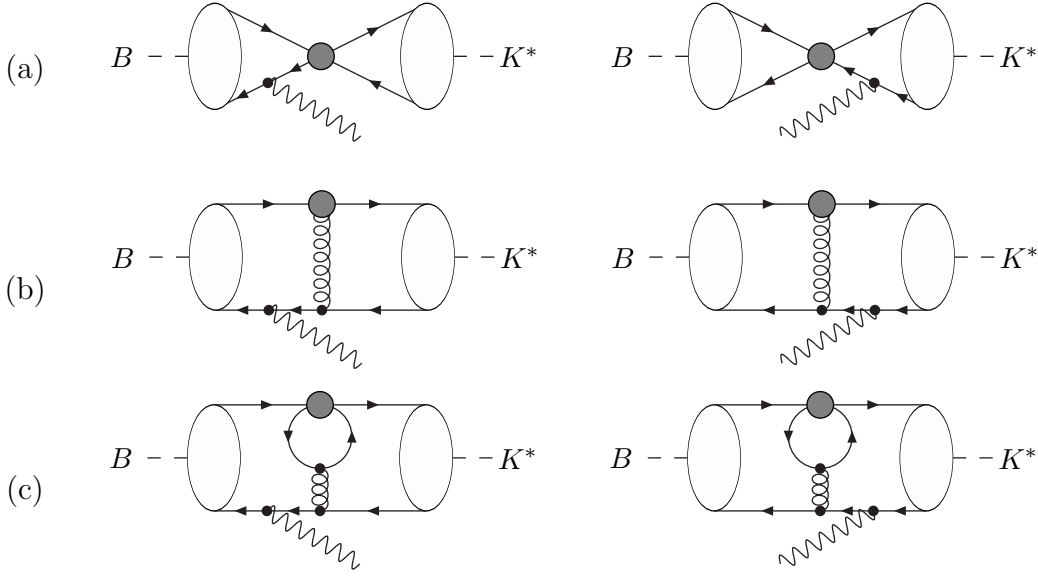

The isospin-asymmetry arises from ``non-factorizable'' graphs where
a photon is radiated from the spectator quark in annihilation or
spectator-scattering diagrams, see Fig.~\ref{dI_fig}.
The corresponding contribution to the
partial decay width depends on
the charge of the spectator quark, and therefore it is different for
charged and neutral $B$~meson decays. The isospin asymmetry vanishes
in naive factorization. Non-factorizable annihilation and
spectator-scattering topologies induce only a tiny
asymmetry when working in the heavy quark limit. However, it has
been observed in~\cite{Kagan:2001zk} that for $B \to K^\ast\gamma$
decays isospin-breaking effects that formally are of sub-leading order in the
$\Lambda_h/m_B$ expansion numerically may be large. Although part of the
calculation requires to model new non-perturbative effects, the
dominant isospin-breaking effect stems from processes which
are dominated by short-distance effects and are thus
perturbatively calculable in the QCD factorization approach.

In the following, we extend the analysis in \cite{Kagan:2001zk} to
the $B \to K^\ast\ell^+\ell^-$ case, including also
longitudinal photon polarizations appearing for $q^2 \neq 0$. 
For that purpose we consider
the exclusive quantities ${\cal C}_9^\perp$ and
${\cal C}_9^\parallel$ introduced in \cite{Beneke:2001at}, and
define
\beq
&&  {\cal C}_9^\perp(q^2) = \left[1+ b_q^\perp(q^2)\right] \, {\cal
  C}_9^{(0)\,\perp}(q^2)
\ , \qquad
 {\cal C}_9^\parallel(q^2) =\left[1+ b_q^\parallel(q^2)\right] \, {\cal
  C}_9^{(0)\,\parallel}(q^2) \ ,
\eeq
where the leading-order expressions 
${\cal C}_9^{(0)\,\perp}(q^2)$ and
${\cal C}_9^{(0)\,\parallel}(q^2)$
are given in (\ref{C9fac}),
and the functions $b_q^\perp(q^2)$ and $b_q^\parallel(q^2)$ 
parameterize non-factorizing effects from photon radiation off
the spectator quark line with flavor $q$. Non-factorizable effects
that are independent of the spectator quark will drop out of the
isospin asymmetry to the considered order of the $\alpha_s$ and
$1/m_b$ expansion.
(Note that there is a formally LO annihilation contribution to ${\cal
  C}_9^{\parallel}$  which does depend on the charge
$e_q$ of the spectator quark and will therefore appear in the
function $b_q^\parallel(q^2)$. Because this contribution is
numerically small we will treat it as a sub-leading effect)
In terms of these new parameters the ($CP$\/--averaged)
isospin asymmetry (\ref{dAI}) can be written as
\beq
  \frac{dA_I}{dq^2} &\simeq&
 {\rm Re}(b_d^\perp - b_u^\perp)
\ \frac{|{\cal C}_9^{(0)\,\perp}|^2}{|{\cal C}_9^{(0)\,\perp}|^2 + |{\cal
    C}_{10}|^2}
\cr && \quad\times
\left(1 +
    \frac14 \cdot \frac{E^2 m_B^2}{q^2 m_{K^\ast}^2} \cdot
    \frac{\xi_\parallel^2(q^2)}{\xi_\perp^2(q^2)} \cdot
     \frac{{\rm Re}(b_d^\parallel-b_u^\parallel) }{ {\rm Re}(b_d^\perp - b_u^\perp)} \cdot
      \frac{|{\cal C}_9^{(0)\,\parallel}|^2}{|{\cal C}_9^{(0)\,\perp}|^2} \right)
\cr
&& \quad \times
\left(1 + \frac14 \cdot  \frac{ E^2 m_B^2}{q^2 m_{K^\ast}^2} \cdot
    \frac{\xi_\parallel^2(q^2)}{\xi_\perp^2(q^2)} \cdot
       \frac{|{\cal C}_9^{(0)\,\parallel}|^2 + |{\cal C}_{10}|^2 }
     {|{\cal C}_9^{(0)\,\perp}|^2+ |{\cal C}_{10}|^2} \right)^{-1}
\label{dAIres}
\eeq
where we neglected terms of order $b_q^2$.
In the limit $q^2 \to 0$ the photon-pole in
${\cal C}_9^{(0)\,\perp}$ will dominate, 
${\cal C}_9^{(0)\,\perp} \sim 1/q^2$, see (\ref{C9fac}).
In this limit the isospin-asymmetry in (\ref{dAIres}) reduces to 
$A_I[B \to K^\ast\gamma]= {\rm  Re}\left[b_d^\perp(0)-b_u^\perp(0)\right]$, 
which is the quantity discussed in \cite{Kagan:2001zk}.
For larger values of $q^2$  the
isospin asymmetry will be dominated by the longitudinal polarization.

Generalizing the definitions in \cite{Kagan:2001zk},
we further define the functions $K_{1,2}^{\perp}$ and $K_{1}^\parallel$ via
\beq
  b_q^\perp(q^2) &=&   \frac{24 \, \pi^2 \, m_B \, f_B \, e_q}{q^2 \,
       \xi_\perp(q^2) \,  {\cal C}_9^{(0)\,\perp}(q^2)}
    \left\{ \frac{f_{K^\ast}^\perp}{m_B} \, K_1^\perp(q^2) +
        \frac{f_{K^\ast} m_{K^\ast}}{6 \lambda_{B,+}(q^2) m_B} \,
        \frac{K_2^\perp(q^2)}{1-q^2/m_B^2}
        \right\} \ , \nonumber \\[0.25em]
  b_q^\parallel(q^2) &=&
    \frac{24\, \pi^2 \,  f_B \, e_q \, m_K^\ast}
     {m_B \, E \, \xi_\parallel(q^2) \, {\cal
        C}_9^{(0)\,\parallel}(q^2)} \left\{ \frac{f_{K^\ast}}{3 \,
      \lambda_{B,-}(q^2)} \, K_1^\parallel(q^2)
 \right\} \ .
\label{bdef}
\eeq
Here
\beq
\lambda_{B,\pm}^{-1}(q^2) &=&
  \int_0^\infty d\omega \, \frac{\phi^B_\pm(\omega)}{\omega-q^2/m_B - i
    \epsilon} \ .
\eeq
are moments of the $B$~meson light-cone distribution amplitudes
as defined for instance in \cite{Grozin:1997pq,Beneke:2000wa}.
The function $K_1^\parallel$ receives an annihilation
contribution to leading order in
the $\Lambda_h/m_B$ expansion with \cite{Beneke:2001at}
\beq
  K_1^\parallel{}^{(a)}(q^2) &=& - \frac{\lambda_u}{\lambda_t} \left(
  \frac{\overline C_1}{3} + \overline C_2 \right) \delta_{qu}
 +  \left( \overline C_4 +  \frac{\overline C_3}{3}\right)
\label{K1parLO}
\eeq
Here the Wilson coefficients $\overline{C}_i$ refer to the
basis defined in~\cite{Beneke:2001at} (at leading order
it coincides with the basis used in \cite{Buchalla:1996vs}).
The functions $K_{1,2}^\perp$ in (\ref{bdef}) only appear
at sub-leading order of the $\Lambda_h/m_B$ expansion. Nevertheless, numerically
the leading order effect in $b_q^\parallel(q^2)$ was found to be
rather small \cite{Beneke:2001at}, whereas the formally sub-leading effects
entering $b_q^\perp(0)$ were found to be sizeable \cite{Kagan:2001zk}.
In particular, the contributions from the scalar penguin
operators $O_5$ and $O_6$ play an important role,
despite the fact that the corresponding Wilson coefficients are
rather small.
In the following we will extend the calculation of Kagan/Neubert in
\cite{Kagan:2001zk} and study how these effects change
with increasing photon virtuality $q^2$.
Note that in this section we will only consider those corrections
that are sensitive to the charge of the spectator quark, and
neglect all sub-leading effects that do {\em not}\/ contribute to
the isospin asymmetry. Furthermore we discard
corrections to $b_q^\parallel$ that are sub-leading in $\Lambda_h/m_B$.
The calculation of such effects
requires a better understanding of the light-cone wave functions
of the $B$~meson
and is beyond the scope of this paper.
However, we have convinced
ourselves that the analogous effects that enhance the isospin
asymmetry for transverse polarizations are absent in the
longitudinal case, see discussion after (\ref{Kgammafac}) below.

\subsubsection{Details of the calculation}

\paragraph{(a) Annihilation graphs:}

The annihilation graphs in Fig.~\ref{dI_fig}(a) involve hadronic
matrix elements of the four-quark
operators ${\cal O}_{1-6}$.
To first approximation, these can be factorized into
two matrix elements defining meson decay constants and
meson-photon transition form factors,
according to the two graphs in Fig.~\ref{dI_fig}(a).
The meson-photon transition form factors  and their role in the
annihilation contribution to $B$~decays have been
extensively discussed in the recent literature
\cite{Descotes-Genon:2002mw}, \cite{Korchemsky:1999qb}-\cite{Melikhov:2001sd}.
%,Khodjamirian:2001ga,Beyer:2001zn,Grinstein:2000pc,Kruger:2002gf,Lunghi:2002ju}.

Let us start with the matrix element of the
operators ${\cal O}_5$ and ${\cal O}_6$
which can be expressed in terms of the (pseudo-)scalar
$K^\ast\gamma^*$ transition form factor and the $B$~meson decay constant $f_B$.
(An analogous term with the (pseudo-)scalar
$B\gamma^*$ transition form factor is absent, 
because the remaining matrix element of
a pseudo-scalar current between a vector meson and the vacuum vanishes.)
Note that the annihilation graphs with ${\cal O}_{5,6}$ are
independent of the $B$~meson wave function at the considered level
of approximation.
We calculate the (pseudo-)scalar $K^\ast\gamma^*$ transition
form factor perturbatively in terms of the $K^\ast$ light-cone wave
functions, which yields
\beq
&&  \langle K^\ast(p',\varepsilon) \gamma^*(q,\mu)
  | \bar s (1+\gamma_5) q | 0\rangle
\nonumber \\[0.1em]
  &=& 2 \, i \, e \, e_q \, f_{K^\ast}^\perp \, \left(
   (q\cdot \varepsilon^*) \,
    p'{}^\mu - \varepsilon^*{}^\mu \, (q \cdot p') 
   - i \epsilon^{\mu\alpha\beta\gamma}\varepsilon^*_\alpha q_\beta
   p'_\gamma \right)
    \int_0^1 du \, \frac{\phi_\perp(u)}{\bar u \, m_B^2 +
      u \, q^2}
\nonumber \\[0.15em]
 &&  \qquad + \left( e_q \to e_s , \ u \to \bar u\right)  \ .
\label{Kgammafac}
\eeq
Note that only transverse polarizations do contribute here.
(This remains true if one includes sub-leading contributions from 
twist-3 wave functions. Therefore the part of the
isospin asymmetry which is related to longitudinal $K^\ast$
polarizations receives no enhanced contributions from
the Wilson coefficients $C_5$ and $C_6$.)
Comparing with (\ref{bdef}) 
we obtain the following contribution to the function $K_1^\perp$
\beq
  K_1^\perp{}^{(a)}(q^2)
 & = & - \left(\overline{C}_6 + \frac{\overline{C}_5}{N_C}\right)
    F_\perp(q^2/m_B^2) \ , \qquad
    F_\perp(\hat s)  = \frac13 \, \int_0^1 du \,
    \frac{\phi_\perp(u)}{\bar u + u \hat s} \ ,
\label{K1perp-O6}
\eeq
which coincides with the result in
\cite{Kagan:2001zk} for $q^2 \to 0$.
(When comparing with \cite{Kagan:2001zk} one has to use
$\lambda_c \approx -\lambda_t$. Note that in (\ref{K1perp-O6})
and in the following we keep terms with $\hat s$ in the denominator of
convolution integrals with light meson distribution amplitudes,
although these terms are formally suppressed by
$\hat s \, \ln \hat s$ for small values of $\hat s = q^2/m_B^2$.)
The functions $K_2^\perp$ and $K_{1}^\parallel$ do not
receive contributions from $C_5$ and $C_6$ to the considered order
in the $\Lambda_h/m_B$ expansion.

The $(V-A)\otimes(V-A)$ currents from the operators 
${\cal O}_1$\/--${\cal O}_4$
give the leading contribution to the function $K_1^\parallel$
quoted in Eq.~(\ref{K1parLO}). For transverse $K^*$ polarizations
one obtains the contribution
\beq
 K_2^\perp{}^{(a)}(q^2) &=& 
- \frac{\lambda_u}{\lambda_t} \left(
  \frac{\overline C_1}{3} + \overline C_2 \right) \delta_{qu}
 +  \left(\overline C_4 +  \frac{\overline C_3}{3} \right) 
\ ,
\label{Bgammacontr}
\eeq
which again coincides with the term quoted in \cite{Kagan:2001zk}.
Note that the separate terms in the decomposition of $\langle
K^\ast\gamma^*|(V-A) \otimes (V-A)|B\rangle$ are invariant
under QED gauge transformations only up to a ``contact term''
which does not depend on the meson's wave functions
\cite{Khodjamirian:2001ga,Beyer:2001zn,Grinstein:2000pc}.
In the sum, of course, these terms cancel. 

%Furthermore the
%perturbative approach to the $K^\ast\gamma^*$ transition form factors,
%taking into account twist-2 and
%twist-3 light-cone distributions amplitudes
%of $K^*$, reproduces the anomalous term discussed in 
%\cite{Beyer:2001zn,Melikhov:2001sd}.

We finally mention that
there is another mechanism to induce the $B \to K^\ast \gamma^{*}$
decay where the photon does not couple in a point-like way, but
enters through its hadronic component \cite{Ali:1995uy,Ball:2002ps}.
This effect gives rise to $1/m_b$ corrections to the meson-photon
transition form factors, which are not taken into account here.
We will also neglect NLO order terms arising from $\alpha_s$ 
corrections to the annihilation
topologies. In the numerical analysis below, we will use leading-order
values for the Wilson coefficients $C_1$ and $C_2$, and
next-to-leading order values for $C_3-C_6$, and include the uncertainty
with respect to variations of the factorization scale $\mu$ into
our error estimate.

\paragraph{(b) Hard spectator interactions involving $O_8$:}

Calculating the decay amplitude from the diagrams in Fig.~\ref{dI_fig}(b),
with the insertion of the gluonic penguin operator ${\cal O}_8$,
we obtain
\beq
 K_1^\perp{}^{(b)}(q^2) &=&
C_8^{\rm eff} \, \frac{m_b}{m_B} \,
\frac{C_F}{N_c} \, \frac{\alpha_s}{4\pi} \,  X_\perp(q^2/m_B^2)
\ ,
\nonumber \\[0.2em]
&& \qquad
X_\perp(\hat s)
=  F_\perp(\hat s) + \frac13 \,
\int_0^1 du \, \frac{1}
                    {(\bar u + u \hat s)^2} \, \phi_\perp(u) \ ,
\label{O8-K1perp}
\\[0.3em]
   K_2^\perp{}^{(b)}(q^2) & = &  {\cal O}(\Lambda_h/m_B) \ ,
\label{O8-K2perp}
\eeq
which again coincides with \cite{Kagan:2001zk} in the limit $q^2 \to
0$. For the longitudinal part we have
\beq
  &&
 K_1^\parallel{}^{(b)}(q^2) =
 - C_8^{\rm eff} \,
  \frac{m_b}{m_B} \,
   \frac{C_F}{N_c} \, \frac{\alpha_s}{4\pi} \, F_\parallel(q^2/m_B^2)
\ , \qquad
F_\parallel(\hat s)
=
 2 \, \int_0^1 du \, \frac{\phi_\parallel(u)}{\bar u + u  \hat s} \ ,
\label{O8-K1par}
\eeq
which has already been calculated in \cite{Beneke:2001at}.
Note that the function $X_\perp^{(1)}(\hat s)$
contains a logarithmic divergence
for $\hat s \to 0$ which indicates the breakdown of QCD
factorization for matrix elements with ${\cal O}_8$
at sub-leading order in the $\Lambda_h/m_B$ expansion.
We follow the authors of \cite{Kagan:2001zk} and parameterize the
divergent part of the integral by replacing
\beq
  \int_0^1 du & \rightarrow & (1+\rho \, e^{i\phi}) \,
  \int_0^{1-\Lambda_h/m_B} du
\eeq
and taking $\Lambda_h \simeq 0.5$~GeV as a regulating cut-off,
together with $0 \leq \rho\leq 1$ and $\phi$, giving a conservative
estimate of the theoretical uncertainty related
to the absolute value and phase
of this non-factorizing soft contribution.

\paragraph{(c) Hard spectator interactions involving ${\cal O}_{1-6}$.}

The calculation of the diagrams in Fig.~\ref{dI_fig}(c) splits into two
parts: First one calculates the internal quark loop which results
in an effective $b \to s g^*$ vertex with vector- and tensor-like
couplings and form factors depending on the internal quark kinematics
and the quark masses in the loop. The tensor contribution is local,
and can be, as usual, absorbed into $C_8^{\rm eff}$. The vector form
factor is expressed in terms of the function ($z\equiv 4m_q^2/s$)
\beq
  h(s,m_q) &=& -\frac{4}{9}\left(\ln\frac{m_q^2}{\mu^2} - \frac{2}{3}
- z \right)- \frac{4}{9} \,(2+z) \,\sqrt{\,|z-1|} \,
\left\{
\begin{array}{l}
\,\arctan\displaystyle{\frac{1}{\sqrt{z-1}}}
\qquad\quad z>1\\[0.4cm]
\,\ln\displaystyle{\frac{1+\sqrt{1-z}}{\sqrt{z}}} - \frac{i\pi}{2}
\quad z\leq 1\end{array}
\right.
\cr &&
\eeq
and reads
\beq
  F_V(s) &=& \frac34 \left\{
  h(s,m_c) \,(\overline{C}_2+\overline{C}_4+\overline{C}_6)
+ h(x,m_b)\,(\overline{C}_3+\overline{C}_4+\overline{C}_6)
\right. \cr  && \qquad  \left.
+ h(s,0)\, (\overline{C}_3+3 \overline{C}_4+3 \overline{C}_6)
-\frac{8}{27}\,(\overline{C}_3-\overline{C}_5-15\overline{C}_6)
\right\} \ .
\eeq
The rest of the calculation is completely analogous
as for the diagrams with ${\cal O}_8$ in Fig.~\ref{dI_fig}(b).
This yields the following contributions to the isospin asymmetry,
\beq
  K_1^\perp{}^{(c)}(q^2) & = &
 \frac{C_F}{N_c} \, \frac{\alpha_s}{4\pi} \,
 \frac23 \, \int_0^1 du \, \frac{\phi_\perp(u)}{ \bar u + u \, \hat
   q^2} \, F_V(\bar u \, m_B^2 + u \, q^2) \ ;
\label{O16-K1perp}
\\[0.3em]
  K_2^\perp{}^{(c)}(q^2) & = &
  - \frac{C_F}{N_c} \, \frac{\alpha_s}{4\pi} \,
    \frac12 \,
    \int_0^1 du \, \left( g_\perp^{(v)}(u) -
      \frac{g_\perp'{}^{(a)}(u)}{4}\right) \, F_V(\bar u \, m_B^2 + u \,
    q^2) \ ;
\eeq
for the transverse part, and
\beq
  K_1^\parallel{}^{(c)}(q^2) & = &
 - \frac{C_F}{N_c} \, \frac{\alpha_s}{4\pi} \,
    2 \, \int_0^1 du \, \phi_\parallel(u) \,  F_V(\bar u \, m_B^2 + u \,
    q^2) \ ;
\label{O16-K1par}
\eeq
for the longitudinal one. In the limit $q^2 \to 0$ we
reproduce the result in \cite{Kagan:2001zk} (neglecting the
small Wilson coefficients $C_{3-6}$ in $F_V(s)$).

\subsubsection{Numerical estimate}

We now turn to the numerical analysis of the isospin asymmetry
in $B \to K^\ast\ell^+\ell^-$, combining the above results as
\beq
  K_1^\perp(q^2) &=&  K_1^\perp{}^{(a)}(q^2) +  K_1^\perp{}^{(b)}(q^2)
  + K_1^\perp{}^{(c)}(q^2) \ ,
\cr
  K_2^\perp(q^2) &=&  K_2^\perp{}^{(a)}(q^2) + K_2^\perp{}^{(c)}(q^2)
\ ,
\cr
  K_1^\parallel(q^2) &=&  K_1^\parallel{}^{(a)}(q^2) +
                          K_1^\parallel{}^{(b)}(q^2) +
                          K_1^\parallel{}^{(c)}(q^2) \ .
\eeq
The central values and uncertainties of all
hadronic and SM  input parameters are taken as
in Ref.~\cite{Beneke:2001at}.
The renormalization scale in the annihilation and hard-scattering
processes will be taken as $\mu' = \sqrt{\mu \Lambda_h}$,
where $\mu$ will be varied as $m_b/2 \leq \mu \leq 2 m_b$.

\begin{figure}[tpbh]
\begin{center}
\epsfclipon
\psfig{bb=180 595 430 750, file=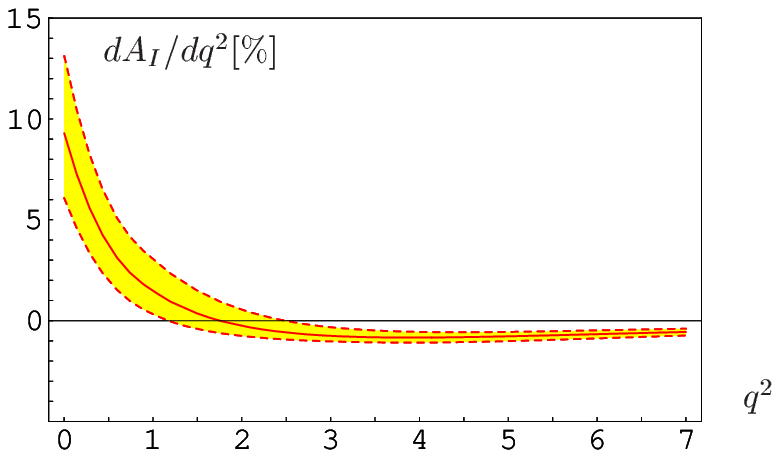, width=0.6\textwidth}
\end{center}
\caption{The SM isospin asymmetry $dA_I/dq^2$
for the decay $B \to K^\ast\ell^+\ell^-$ as a function of $q^2$.
The error band includes the variation of hadronic and
SM  input parameters (taken as in \cite{Beneke:2001at})
and of the renormalization scale $\mu$ between $m_b/2$ and $2 m_b$.}
\label{dAIfig}
\end{figure}

The result for the isospin asymmetry is plotted as a function
of $q^2$ in Fig.~\ref{dAIfig}.
At $q^2=0$ we basically recover the result of Kagan/Neubert
\cite{Kagan:2001zk} with
\beq
  {\rm Re}\left[{b_d^\perp(0)-b_u^\perp(0)}\right] &=& + \, 9.3
  {\,}^{+3.8}_{-3.2} \, \% \ ,
\eeq
where the uncertainty due to the theoretical input parameters
(taken as in  \cite{Beneke:2001at}) is dominated by the
soft form factor $\xi_\perp$ ($\sim {}^{+2.3}_{-1.5}\%$), the
$B$~meson decay constant ($\sim 1.5\%$), the moment of the
$B$~meson light-cone wave function $\lambda_B$ ($\sim 1.1\%$),
and the IR-sensitive integral $X_\perp^{(1)}$ ($\sim 1\%$).
We thus confirm the conclusion in \cite{Kagan:2001zk} that QCD factorization
correctly reproduces the sign and magnitude of the experimentally
measured isospin asymmetry ($A_I[B \to K^\ast\gamma] = 0.11 \pm 0.07$
\cite{Coan:1999kh,Ushiroda:2001sb,Nash:2002vr}) with SM
values for the Wilson coefficients.

For increasing values of $q^2$ the isospin-asymmetry decreases,
and its central value becomes slightly negative above
$q^2 = 2$~GeV$^2$ and stays basically at a constant value
of about $-1\%$. Since the uncertainty related to the hadronic input
parameters is reduced as well, 
this means that the measurement of a significant
deviation from zero of the isospin asymmetry in the range
$2~{\rm GeV}^2 < q^2 < 7~{\rm GeV}^2$ may still indicate
new physics (although one would need to have a handle on
even higher order effects, before drawing any definite
conclusions). Note that in the SM 
the isospin asymmetry is sensitive to $C_5$ and $C_6$
at small $q^2$ but to $C_3$ and $C_4$ at larger $q^2$.
Thus, in principle, the two momentum regions provide complementary tests
of the four-quark penguin-operators (see also Fig.~\ref{dAIfig2} below).

%%%%%%%%%%%%%%%%%%%%%%%%%%%%%%%%%%%%%%%%%%%%%%%%%%%%%%%%%%%%%%%%%%%%%%%%%%

\section{Decay asymmetries in MSSM with MFV}

In this section we present an example to illustrate
how new physics may affect the FB and isospin asymmetry in 
$B \to K^*\ell^+\ell^-$ decay. 
We consider the minimal supersymmetric standard model (MSSM)
with minimal flavor violation (MFV) \cite{Ciuchini:1998xy,list,amb}. 
In this model, all flavour transitions are governed by the CKM matrix.
In addition, in order to increase the predictivity of the
model, we impose the following restrictions.
\begin{itemize}
\item All  supersymmetric particles  are taken as heavy (about 1~TeV),
except for charginos, sneutrinos, the light (mostly right-handed) stop, and
charged Higgs fields.
Heavy particles (squarks, gluinos etc.) are integrated out to
obtain a ``low-energy'' effective theory in terms of light SUSY particles and
SM fields~\cite{Ciuchini:1998xy}.
\item The weak effective Hamiltonian (\ref{heff}) includes only SM operators.
\item We assume a flavor diagonal mass matrix 
in the  down-squark sector \cite{Bobeth:1999ww}.
Therefore neutralino contributions to the Wilson coefficients will
be absent.
\end{itemize}
We  carefully consider the two regimes of small and large $\tan\beta$,
taking into account the modifications and subtleties associated
with large values of $\tan\beta$ \cite{Degrassi:2000qf,private} 
($\tan\beta$ is the ratio of vacuum expectation values of
the two Higgs doublets).

The analysis is organized in three steps:
First, we  perform a scan of model parameters within 
specific ranges which will be indicated below. 
For each set of parameter values we will calculate
the Wilson coefficients contributing to the $B \to X_s\gamma$ decay
at NLO following 
~\cite{Ciuchini:1998xy,Degrassi:2000qf,private},
including also electroweak corrections~\cite{Czarnecki:1998tn}
and non-perturbative contributions~\cite{Buchalla:1998ky}.
Then we will use the experimental data on $B
\to X_s \gamma$ to constrain the model parameter space, i.e.\
those parameter sets that lead to Wilson coefficients which
are incompatible with the $B \to X_s \gamma$ constraint will be 
discarded.
(Note that the allowed region of parameter space will divide into
two classes, corresponding to negative or positive sign 
of $C_7^{\rm eff}$.)
Finally, using the remaining part of parameter space we evaluate 
the Wilson coefficients in (\ref{heff}), 
in terms of which we obtain numerical results for the two decay asymmetries, 
$dA_{\rm FB}$ and $dA_I$ in MFV.

\subsection{Model set-up}

\label{mfvsetup}

The free parameters of the model are:
\begin{itemize}

\item The Higgs mixing parameter $\mu$. We consider
       $|\mu| \leq 500$ GeV.

\item The mass parameter $M_2$ in the chargino mass matrix
(``weak gaugino mass''), 
which is varied as $M_2=100-800$ GeV.
Alternatively, one could consider one of the chargino mass eigenstates.
In the parameter scan we always impose the constraint that
the masses of both charginos should be heavier than $100~{\rm GeV}$,
as required by direct searches.

\item The light stop mass is taken in the range
$100 \leq {\tilde m}_{t2} \leq 350$ GeV. 
(Note that depending on the relative size of stop, gluino and other
squark masses, one has to be careful when evaluating the 
Wilson coefficients. In our case, we are
considering a scenario with a very light right-handed stop and charginos 
in comparison with gluino and other squark masses. This implies that
we have to use the Wilson coefficients at the weak scale as 
described in Eq.(31) of \cite{Degrassi:2000qf}.)

\item $|\theta_{t}|$ is the stop mixing angle, such that 
${\tilde t}_1={\rm cos} \theta_t {\tilde t}_L+
{\rm sin}\theta_t {\tilde
t}_R$ and $
{\tilde t}_2=-{\rm sin} \theta_t {\tilde t}_L + {\rm cos}\theta_t {\tilde 
t}_R$. 
Constraints from electroweak precision measurements
favor the light stop quark to be almost right-handed~\cite{Altarelli:1997et}.
We therefore take  $|\theta_{t}| \leq \pi/10$.

\item The charged Higgs mass is considered in the range
  $M_H=100-400$~GeV. We have chosen a not too heavy mass for
the charged Higgs to evaluate the maximal possible impact in our
observables of a non-decoupled charged Higgs.

\item The ratio of vacuum expectation values for the two Higgs
  doublets is varied as $\tan \beta=2-40$. 
In the expressions for the Wilson Coefficients
the large $\tan \beta$ regime is treated as in 
Ref.~\cite{Degrassi:2000qf}. 
Large values of $\tan \beta$ are 
particularly interesting since this can easily lead to a sign flip 
in the value of $C_7^{\rm eff}$.
\end{itemize}
The following parameters will be kept
fixed in the analysis:
\begin{itemize}
\item[-]
According to the 
effective theory approximation of the MFV supersymmetric model,
all heavy squark masses (including the heavy stop) are taken at
a common value, $m_{\tilde q}$=1~TeV.
%\item[-]
Also the gluino mass is taken at a fixed value $m_{\tilde g}$=1~TeV. 
The renormalization effects from integrating out  
squarks and gluinos between the scale 1~TeV and the electroweak
scale as discussed in \cite{Degrassi:2000qf} are taken into account.

\item[-] For the sneutrino masses we consider a 
conservative value, $m_{\tilde \nu}$=100~GeV which is
not in conflict with direct searches.

\end{itemize}
Scanning over discrete points in the so-defined parameters space,
we calculate the Wilson coefficients for the operators in (\ref{heff})
at the electroweak scale, $\mu_W = {\cal O}(M_W)$. 
At NLO accuracy one obtains the generic form
\begin{eqnarray}
C_i(\mu_W) &=& C_i(\mu_W)_{\rm SM} + \delta C_i(\mu_W)
\nonumber \\[0.15em]
&=&C_i^{(0)}(\mu_W)_{SM}+ \delta
C_i^{(0)}(\mu_W)_H+\delta C_i^{(0)}(\mu_W)_{\rm SUSY}
\nonumber \\ && {} + {\alpha_s(\mu_W) \over 4 \pi} \, 
\left(C_i^{(1)} (\mu_W)_{SM}+\delta C_i^{(1)}(\mu_W)_H+
\delta C_i^{(1)}(\mu_W)_{\rm SUSY} \right)\, ,
\label{deltadef}
\end{eqnarray}
where we indicated explicitely the contributions from different
virtual particles in box and penguin diagrams:
\begin{itemize}
\item SM contributions from charged gauge bosons and up-type quarks.
\item Contributions from charged Higgs bosons and up-type quarks,
 present in SUSY, but also in generic models with two Higgs doublets.
\item Specific SUSY contributions from charginos and up-type squarks,
gluinos and down-type squarks, neutralinos and down-type quarks.
(We will
mainly focus on chargino effects. In our model gluinos
are very heavy and their main effect is to induce 
renormalization effects in the vertices involving charginos, charged 
Higgs bosons, charged gauge bosons and unphysical scalars 
\cite{Ciuchini:1998xy}. As stated above, neutralino contributions
are absent due to our assumption about the mass matrix in the 
down-quark sector.)
\end{itemize}
The expressions for $\delta C_{7,8}^{\rm eff}{}^{(0,1)}$ 
can be found in Ref~\cite{Ciuchini:1998xy};
SUSY contributions to $C_9$ and $C_{10}$ are taken 
from~\cite{Cho:1996we,Hewett:1997ct}.
We have also included the NLO SUSY contribution to the 
four-quark penguin operators from Ref.~\cite{Bobeth:1999ww}. 
(Notice the different conventions used for $\theta_t$ and the sign of 
the $\mu$ parameter in~\cite{Ciuchini:1998xy} and
~\cite{Bobeth:1999ww}.)
The values of the Wilson coefficients are evolved to 
the low scales relevant for the exclusive $B$ decays by
standard renormalization group analysis.

\subsection{Constraints from  $ B \to X_s \gamma$}

We consider the world average of measurements of the
inclusive branching ratio for $B \rightarrow X_s \gamma$~\cite{expbsgamma}
\beq
&&
2.82 \times 10^{-4} \leq {\rm BR}(B \to X_s \gamma) \leq 3.62
\times 10^{-4} \, .
\eeq
The precision reached in the determination of
this decay mode significantly constrains new physics contributions
and requires to consider very carefully higher order QCD effects
in the theoretical calculations, also in the
presence of new physics. For instance, at two loops the electroweak 
amplitude gets enhanced in the SM due to a large QCD mixing with the 
effective operator $b \to s {\bar c} c$. 
Moreover, in the MFV model we are discussing it is known that there is a
compensatory effect between the charged Higgs and chargino-stop
contribution that gets unbalanced once the NLO correction is
included. 
Apart from the standard renormalization group improvement of
Wilson coefficients at NLO, we finally
have to include non-perturbative corrections associated with
the hadronic matrix elements for $B \to X_s\gamma$.
In the analysis we will strictly follow 
Refs.~\cite{Ciuchini:1998xy,Degrassi:2000qf}.

Scanning over the allowed parameter space
(notice that in our scenario we are only exploring 
light stop, charginos and not too heavy charged Higgses),
we find the following allowed range for the new
contributions $\delta C_7$ to the Wilson coefficient 
$C_7^{\rm eff}$ at the scale $\mu_W$ (see Eq.~(\ref{deltadef}))
\beq
 -0.02 \leq \delta C_7(\mu_W) \leq 0.12 \quad \mbox{or} \quad
  0.59 \leq \delta C_7(\mu_W) \leq 1.24 \, ,
\label{C7range}
\eeq
where the first range is SM-like (negative sign for $C_7^{\rm eff}$),
and the second range corresponds to the flipped-sign solution.
The behaviour of the two ranges as a function of the 
parameters $\tan \beta$, $\theta_t$ and ${\tilde m}_{t2}$ is quite
different. We have found
that the points in the SM-like region corresponds predominantly to values for
$\tan \beta$ in the low regime (between 2 to 10). Moreover, higher values
of $m_h$ tend to prefer lower values of $\tan \beta$. There is no preferred
value for $\theta_t$ and also any value of ${\tilde m}_{t2}$ is found.
The flipped-sign solutions for $C_7^{\rm eff}$ correspond to
large values of $\tan \beta$ of order $40$. In this case, $\theta_t$
preferably takes extreme values around $\pm \pi/10$.
The allowed values for ${\tilde m}_{t2}$ increase with 
increasing $m_h$ but are always below 300~GeV, 
for example, for $m_h=100$~GeV one gets ${\tilde m}_{t2}=100-240$~GeV,
and for the highest value $m_h=400$~GeV one has ${\tilde m}_{t2}=100-300$~GeV.
The corresponding range for the Wilson coefficient $C_8^{\rm eff}$ reads
\beq
 -0.18 \leq \delta C_8(\mu_W) \leq 0.03 \quad \mbox{or} \quad
 0.20 \leq \delta C_8(\mu_W) \leq 0.57 \, ,
\label{C8range}
\eeq

Concerning the Wilson coefficients $C_3-C_6$ of four-quark penguin
operators our MFV model is SM-like. 
Also the Wilson coefficients $C_9$ and $C_{10}$ only receive
tiny contributions (of at most 10\% from their SM value, we restricted
ourselves to LO SUSY contributions here). 
The mechanisms that enhance the SUSY contributions to 
$C_7^{\rm eff}$ at large $\tan \beta$ are not working for $C_9$ and 
$C_{10}$. For instance, the charged Higgs contribution dominant for
$C_7^{\rm eff}$ at large $\tan\beta$ is suppressed for $C_{9,10}$
\cite{Cho:1996we,Hewett:1997ct}. 
Therefore the modifications 
for the FB and isospin-asymmetry, to be discussed below, 
will be due to new physics 
contributions to $C_7^{\rm eff}$, $C_8^{\rm eff}$ mainly.

\subsection{FB and isospin asymmetry in MFV}

\begin{figure}[thpb]
\begin{center}
  \psfig{file=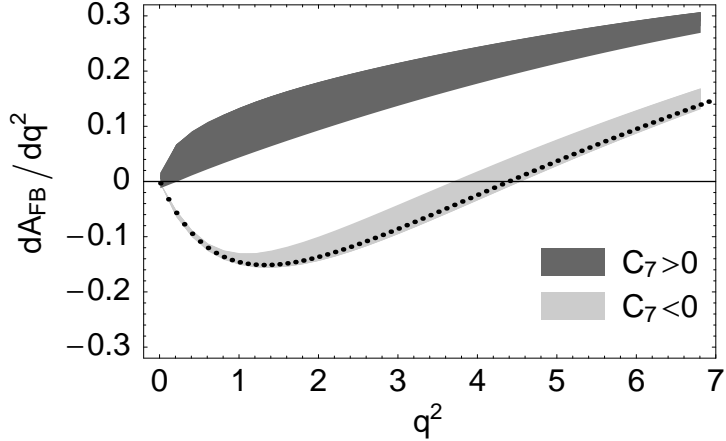,width=0.6\textwidth}
\end{center}
\caption{FB asymmetry (\ref{dAFB}) for the decay $B
\to K^*\ell^+\ell^-$ as a function of $q^2$. The light band
corresponds to the region of parameter space with same (negative) sign for
$C_7$ as in the SM; the dark band refers to solutions with flipped
sign for $C_7$; and the dotted line is the central value
of the SM prediction.} 
\label{fbsusy}
\end{figure}

Using the parameter values in the MFV scenario
allowed from the constraints discussed
in the previous subsection, we calculate 
the FB and isospin-asymmetry as explained in section~\ref{sec2}.
The result for the FB asymmetry is shown in Fig.~\ref{fbsusy}.
The two distinct bands refer to the two possible
signs of $C_7^{\rm eff}$ in (\ref{C7range}).
Comparing with the SM case in Fig.~\ref{fig:dafbnorm},
we observe that the variation of MFV parameters for
the case of SM-like sign of $C_7^{\rm eff}$ (light band) is
indistinguishable from the hadronic uncertainties.
An exception is the FB asymmetry zero where the hadronic
uncertainty is reduced due to the form factor relations
discussed in \cite{Burdman:1998mk,Ali:1999mm,Beneke:2001at}.
Here the MFV effects are competitive with the hadronic uncertainty,
and show a tendency towards lower values of the FB asymmetry zero.
This implies that if the experimental measurement of the FB asymmetry
zero found an even {\em larger}\/ values than predicted in the SM
it would  not easily be accommodated by our MFV scenario, and
eventually would lead to the exclusion of the region of parameter
space described in section~\ref{mfvsetup}.
As is well known, for the flipped sign solution of $C_7^{\rm eff}$ (dark band)
the characteristic features of the FB asymmetry change dramatically:
$dA_{FB}/dq^2$ is positive for small {\em and}\/ large values of $q^2$, 
and there is no asymmetry zero in the spectrum.

\begin{figure}[ptbh]
\begin{center}
  \psfig{file=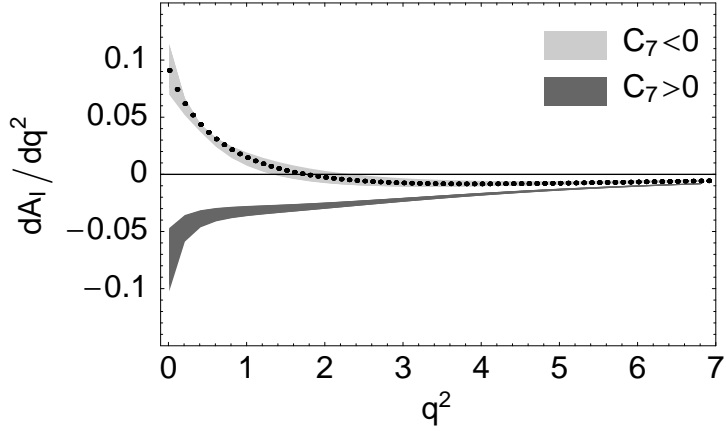,width=0.6\textwidth}
\end{center}
\caption{Isospin asymmetry (\ref{dAI}) for the decay $B \to
K^*\ell^+\ell^-$ as a function of $q^2$. Notations as in
Fig.~\protect\ref{fbsusy}.}
\label{isosusy}
\end{figure}

The MFV prediction for the isospin asymmetry, shown in Fig.~\ref{isosusy}, 
also splits into two bands, according to the sign of $C_7^{\rm eff}$. 
Because the isospin asymmetry is dominated by the penguin coefficients
$C_{3-6}$ which are only slightly modified in our MFV scenario,
both bands are very thin. Within the MFV the isospin asymmetry at
small values of $q^2$ thus provides a possibility to exclude the flipped sign
solution for $C_7^{\rm eff}$, independently from the analysis of the FB
asymmetry (as already mentioned, at present, experimental data favor 
the same-sign solution). On the other hand, a measurement of a large
isospin asymmetry at moderate values of $q^2$  would point to physics
beyond both, the SM and MFV.

\section{Beyond MFV}

As we have seen in the previous section, in MFV the main effect
on the FB and isospin asymmetry is due to a possible
sign flip in the value of $C_7^{\rm eff}$. Taking the central value
of the experimental measurement of the isospin asymmetry in $B \to
K^*\gamma$ at face value, one is tempted to consider this
scenario as already excluded. (A reduction 
of theoretical and experimental uncertainties for this observable, 
combined with a measurement of the FB asymmetry 
would allow for an ultimate conclusion.)
The  Wilson coefficients for semi-leptonic and 4-quark penguin operators
are only slightly affected 
in our MFV scenario, and therefore for SM-like sign of $C_7^{\rm eff}$
MFV is indistinguishable from the SM within the theoretical
uncertainties. 

On the other hand,  in a generic new physics model, one might expect
sizeable corrections to other Wilson coefficients as well, which may
show up in the $B \to K^*\ell^+\ell^-$ decay asymmetries.
In the following we perform a brief 
discussion on the possible size of such effects.

\begin{figure}[tpb]
\begin{center}
\epsfclipon
(a)
\psfig{bb=180 595 410 735, file=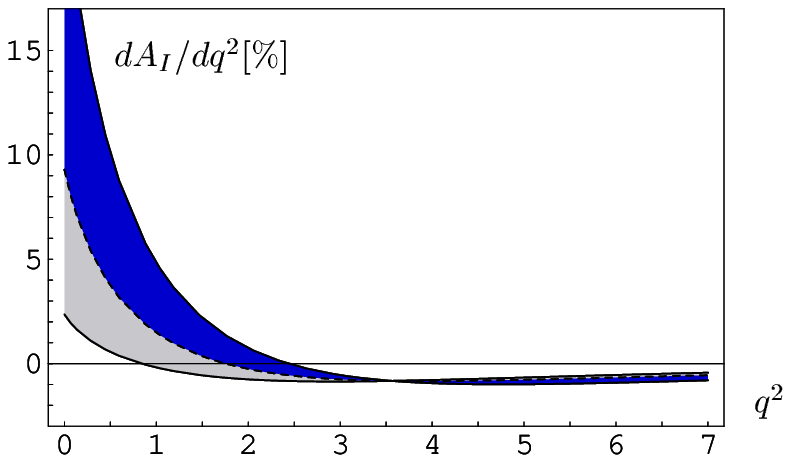, width=0.44\textwidth}
\hskip0.5em
(b)
\psfig{bb=180 595 410 735, file=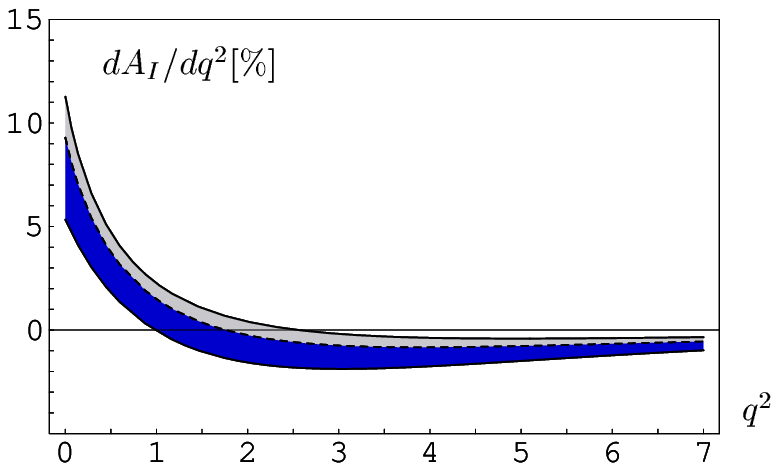, width=0.44\textwidth}
\end{center}
\caption{The isospin asymmetry $dA_I/dq^2$
for the decay $B \to K^\ast\ell^+\ell^-$ as a function of $q^2$.
(a) The combination $a_6^{(0)} = 
(\bar C_6+\bar C_5/3)$ in the function $K_1^{\perp(a)}$
is varied within a factor of two around its SM value
(dark band = larger values, light band = smaller values).
(b) The same for the combination $a_4^{(0)}=(\bar C_4+\bar C_3/3)$ in 
the functions $K_1^{\parallel(a)}$ and $K_2^{\perp(a)}$.
}
\label{dAIfig2}
\end{figure}

\begin{itemize}
\item
The penguin operators ${\cal O}_{3-6}$ give the main effect to
the isospin asymmetry in $B \to K^*\gamma$ and $B \to
K^*\ell^+\ell^-$. More precisely, $dA_I/dq^2$ is sensitive to the
combinations $a_6^{(0)}=(\bar C_6+ \bar
C_5/3)$ and  $a_4^{(0)}=(\bar C_4 + \bar C_3/3)$ entering the
functions $K_{1,2}^{\perp(a)}$ and $K_1^{\parallel(a)}$ that describe
the dominant annihilation contribution to the isospin asymmetry
at small and large $q^2$, respectively.
In principle, the size of the
corresponding Wilson coefficients is already
constrained from many sources, for instance non-leptonic $B$ decays,
life-time differences of $B$~mesons, kaon decays etc.
Since the sensitivity to individual Wilson coefficients
in these decays is not very high,
and considering the possibility of some conspiracy between different
new physics effects one may, however, still allow for
sizeable deviations of $C_{3-6}$ from their SM values.
Fig.~\ref{dAIfig2} shows the effect on the isospin asymmetry
if we allow for a variation of $a_6^{(0)}$
and $a_4^{(0)}$ within a factor of two. In this case,
as already mentioned above, modifications of
$a_6^{(0)}$ (respectively $a_4^{(0)}$)
are competitive with the theoretical uncertainties at
small (respectively large) values of $q^2$ (at large $q^2$ values, however,
the isospin asymmetry itself is rather small, and probably the
necessary accuracy to see modifications of $a_4^{(0)}$ will be hard
to achieve.)

\item
The semi-leptonic operators ${\cal O}_{9,10}$ are relevant for
$B \to K^*\ell^+\ell^-$ at not too small values of $q^2$.
As already stated above the FB asymmetry provides
a rather clean test of the Wilson coefficients $C_{9,10}$
(compared to this the isospin asymmetry is less sensitive
to $C_9$ and $C_{10}$, unless the experimental measurement
leads to an unexpectedly large asymmetry for $2$~GeV$^2<q^2<7$~GeV$^2$.)
In particular, when NLO QCD effects are included the Wilson
coefficient $C_9$ can be measured with 10\% accuracy from the
FB asymmetry zero \cite{Beneke:2001at} (provided
the relative signs of $C_{7,9,10}$ are SM-like). 
If the relative sign between $C_7^{\rm eff}$ and $C_9$ is flipped
there will be no FB asymmetry zero at all; in case of a non-SM-like
sign of $C_{10}$ the overall sign of the FB asymmetry will change.
We refrain from giving quantitative results here, but rather
refer to \cite{Ali:2002jg,Ali:1999mm,Safir:2002gc} 
for extensive phenomenological analyses of these issues.

\item 
In this work we do not entertain the possibility of introducing new operators 
in the effective Hamiltonian~(\ref{heff}). 
An example of contributions of non-standard operators to the
FB asymmetry in $B \to K^*\ell^+\ell^-$ can be found in
\cite{Buchalla:2000sk}. A discussion of non-standard penguin
operators, which would also be relevant for the isospin asymmetry,
has been given in \cite{Grossman:1999av}.

\end{itemize}

\section{Summary}

We discussed two dedicated observables in $B \to K^*\ell^+\ell^-$
decays, the forward-backward and isospin-asymmetry, which can be used
to test the standard model against new physics. We concentrated on the
large recoil region where the decay amplitudes can be calculated using
the QCD factorization approach. (For the FB asymmetry we used the
results from \cite{Beneke:2001at}; for the isospin asymmetry we
generalized the $B \to K^*\gamma$ analysis performed in
\cite{Kagan:2001zk}.) The two asymmetries turn out to give
complementary tests of the Wilson coefficients in the weak
effective Hamiltonian: For a given value (and sign) of $C_7^{\rm
  eff}$, the FB~asymmetry is sensitive to the
semi-leptonic coefficients $C_9$ and $C_{10}$. On the other hand
the isospin asymmetry is dominated by the effects of four-quark
coefficients $C_{5,6}$ and $C_{3,4}$ at small or medium values of
$q^2$.

To illustrate possible new physics effects we considered an
explicit model, namely minimal supersymmetry with minimal flavor
violation. In this scenario the main deviation from the SM is the
possibility to generate another sign for the magnetic penguin
coefficients $C_7^{\rm eff}$, $C_8^{\rm eff}$ 
(while the absolute value for $C_7^{\rm eff}$ is constrained
by the $B \to X_s\gamma$ measurement). The corresponding two regions
of allowed  parameter sets translate into two well separated bands
of predictions for the FB and isospin-asymmetry. On the other hand,
the change in the remaining Wilson coefficients is rather small in the
MFV scenario and indistinguishable from hadronic uncertainties of the
theoretical calculation.

Considering an alternative point of view and assuming the SM to be a
valid approximation in rare $B$ decays, 
the simultaneous study of the isospin asymmetry and the
forward-backward asymmetry in the large-recoil region
($q^2 < 4 m_c^2$) provides a perfect testing ground for the
QCD factorization approach.

\section{Acknowledgements}

T.F. and J.M. likes to thank Antonio Masiero, Martin Beneke, Giuseppe Degrassi, 
Paolo Gambino, Dimitri Melikhov and Dirk Seidel for helpful
discussions. J.M. acknowledges financial support by CICYT Research Project 
AEN99-0766, FPA2002-00748 and the Theory Division at CERN.


\begin{thebibliography}{99}


\bibitem{smbsg}
B.~Grinstein, M.~J.~Savage and M.~B.~Wise,
%``B $\to$ X(S) E+ E- In The Six Quark Model,''
Nucl.\ Phys.\ B {\bf 319}, 271 (1989);
%%CITATION = NUPHA,B319,271;%%
%\bibitem{Buras:xp}
A.~Ali and C.~Greub,
%``Inclusive Photon Energy Spectrum In Rare B Decays,''
Z.\ Phys.\ C {\bf 49}, 431 (1991);
%%CITATION = ZEPYA,C49,431;%%;
%``A Profile Of The Final States In B $\to$ X(S) Gamma And An Estimate Of The
%Branching Ratio Br (B $\to$ K* Gamma),''
Phys.\ Lett.\ B {\bf 259}, 182 (1991);
%%CITATION = PHLTA,B259,182;%%;
Phys.\ Lett.\ B {\bf 361}, 146 (1995).
%%CITATION = HEP-PH 9506374;%%;
A.~J.~Buras, M.~Misiak, M.~M{\"u}nz and S.~Pokorski,
%``Theoretical Uncertainties And Phenomenological Aspects Of B $\to$ X(S) Gamma 
%Decay,''
Nucl.\ Phys.\ B {\bf 424}, 374 (1994)
%%CITATION = HEP-PH 9311345;%%;
%\bibitem{Ciuchini:1994xa}
M.~Ciuchini, E.~Franco, G.~Martinelli, L.~Reina and L.~Silvestrini,
%``b $\to$ s gamma and b $\to$ s g: A Theoretical reappraisal,''
Phys.\ Lett.\ B {\bf 334}, 137 (1994).
%%CITATION = HEP-PH 9406239;%%;
%\bibitem{Adel:1993ah}
K.~Adel and Y.~P.~Yao,
%``Exact alpha-s calculation of b $\to$ s + gamma b $\to$ s + g,''
Phys.\ Rev.\ D {\bf 49}, 4945 (1994).
%%CITATION = HEP-PH 9308349;%%;
C.~Greub, T.~Hurth and D.~Wyler,
%``Virtual $O(\a_s)$ corrections to the inclusive decay $b \to s
%\gamma$,''
Phys.\ Rev.\ D {\bf 54}, 3350 (1996).
%%CITATION = HEP-PH 9603404;%%
%\bibitem{Ali:1990tj}
C.~Greub, T.~Hurth and D.~Wyler,
%``Virtual corrections to the decay $b \to s \gamma$,''
Phys.\ Lett.\ B {\bf 380}, 385 (1996).
%%CITATION = HEP-PH 9602281;%%;
K.~G.~Chetyrkin, M.~Misiak and M.~M{\"u}nz,
%``Weak radiative B-meson decay beyond leading logarithms,''
Phys.\ Lett.\ B {\bf 400}, 206 (1997)
[Erratum-ibid.\ B {\bf 425}, 414 (1998)].
%%CITATION = HEP-PH 9612313;%%;
A.~J.~Buras, A.~Kwiatkowski and N.~Pott,
%``Next-to-leading order matching for the magnetic photon penguin operator  in 
%the B $\to$ X/s gamma decay,''
Nucl.\ Phys.\ B {\bf 517}, 353 (1998).
%%CITATION = HEP-PH 9710336;%%
P.~Ball {\it et al.},
%``B decays at the LHC,''
hep-ph/0003238.
%%CITATION = HEP-PH 0003238;%%
P.~Gambino and M.~Misiak,
%``Quark mass effects in anti-B $\to$ X/s gamma,''
Nucl.\ Phys.\ B {\bf 611} (2001) 338.
%%CITATION = HEP-PH 0104034;%%


\bibitem{Ciuchini:1998xe}
M.~Ciuchini, G.~Degrassi, P.~Gambino, and G.~F. Giudice,
\newblock Nucl. Phys. {\bf B527}, 21 (1998), hep-ph/9710335.
%%CITATION = HEP-PH 9710335;%%


\bibitem{Czarnecki:1998tn}
A.~Czarnecki and W.~J.~Marciano,
%``Electroweak radiative corrections to b $\to$ s gamma,''
Phys.\ Rev.\ Lett.\  {\bf 81}, 277 (1998)
[hep-ph/9804252].
%%CITATION = HEP-PH 9804252;%%
P.~Gambino and U.~Haisch,
%``Complete electroweak matching for radiative B decays,''
JHEP {\bf 0110}, 020 (2001)
[hep-ph/0109058].
%%CITATION = HEP-PH 0109058;%% 


\bibitem{Buchalla:1996vs}
G.~Buchalla, A.~J. Buras, and M.~E. Lautenbacher,
\newblock Rev. Mod. Phys. {\bf 68} (1996) 1125 [hep-ph/9512380].
%%CITATION = HEP-PH 9512380;%%

\bibitem{Buchalla:1998ky}
G.~Buchalla, G.~Isidori and S.~J.~Rey,
%``Corrections of order Lambda(QCD)**2/m(c)**2 to inclusive rare B
%decays,''
Nucl.\ Phys.\ B {\bf 511}, 594 (1998)
[hep-ph/9705253].
%%CITATION = HEP-PH 9705253;%%


\bibitem{smbsll}
M.~Misiak,
%``The B $\to$ S E+ E- And B $\to$ S Gamma Decays With Next-To-Leading 
%Logarithmic QCD Corrections,''
Nucl.\ Phys.\ B {\bf 393}, 23 (1993);
[Erratum-ibid.\ B {\bf 439}, 461 (1995)];
A.~J.~Buras and M.~M{\"u}nz,
%``Effective Hamiltonian for B $\to$ X(s) e+ e- beyond leading logarithms in 
%the NDR and HV schemes,''
Phys.\ Rev.\ D {\bf 52}, 186 (1995).
%%CITATION = HEP-PH 9501281;%%;
C.~Bobeth, M.~Misiak and J.~Urban,
%``Photonic penguins at two loops and m(t)-dependence of BR(B $\to$ X(s) l+  
%l-),''
Nucl.\ Phys.\ B {\bf 574}, 291 (2000)
[hep-ph/9910220].
%%CITATION = HEP-PH 9910220;%%
H.~H.~Asatrian, H.~M.~Asatrian, C.~Greub and M.~Walker,
%``Two-loop virtual corrections to B $\to$ X/s l+ l- in the standard model,''
Phys.\ Lett.\ B {\bf 507}, 162 (2001).
%%CITATION = HEP-PH 0103087;%%;
H.~H.~Asatryan, H.~M.~Asatrian, C.~Greub and M.~Walker,
%``Calculation of two loop virtual corrections to b $\to$ s l+ l- in the  
%standard model,''
Phys.\ Rev.\ D {\bf 65}, 074004 (2002).
%%CITATION = HEP-PH 0109140;%%
H.~M.~Asatrian, K.~Bieri, C.~Greub and A.~Hovhannisyan,
%``NNLL corrections to the angular distribution and to the  forward-backward 
%asymmetries in b $\to$ X/s l+ l-,''
hep-ph/0209006.
%%CITATION = HEP-PH 0209006;%%
A.~Ghinculov, T.~Hurth, G.~Isidori and Y.~P.~Yao,
%``Forward-backward asymmetry in B $\to$ X/s l+ l- at the NNLL level,''
hep-ph/0208088;
%%CITATION = HEP-PH 0208088;%%;
hep-ph/0211197.
%%CITATION = HEP-PH 0211197;%%


\bibitem{bbmr}
S.~Bertolini, F.~Borzumati, A.~Masiero and G.~Ridolfi,
%``Effects Of Supergravity Induced Electroweak Breaking On Rare B Decays And 
%Mixings,''
Nucl.\ Phys.\ B {\bf 353}, 591 (1991).
%%CITATION = NUPHA,B353,591;%%


\bibitem{susybsg}          
R.~Barbieri and G.~F.~Giudice,
%``b $\to$ s gamma decay and supersymmetry,''
Phys.\ Lett.\ B {\bf 309}, 86 (1993)
[hep-ph/9303270].
%%CITATION = HEP-PH 9303270;%%
F.~M.~Borzumati,
%``The Decay b $\to$ s gamma in the MSSM revisited,''
Z.\ Phys.\ C {\bf 63}, 291 (1994)
[hep-ph/9310212].
%%CITATION = HEP-PH 9310212;%%


\bibitem{Bobeth:1999ww}
C.~Bobeth, M.~Misiak and J.~Urban,
%``Matching conditions for b $\to$ s gamma and b $\to$ s gluon in
%extensions  of the standard model,''
Nucl.\ Phys.\ B {\bf 567}, 153 (2000)
[hep-ph/9904413].
%%CITATION = HEP-PH 9904413;%%





\bibitem{Ciuchini:1998xy}
M.~Ciuchini, G.~Degrassi, P.~Gambino and G.~F.~Giudice,
%``Next-to-leading {QCD} corrections to B $\to$ X/s gamma in
%supersymmetry,''
Nucl.\ Phys.\ B {\bf 534}, 3 (1998)
[hep-ph/9806308].
%%CITATION = HEP-PH 9806308;%%
 



%\cite{Degrassi:2000qf}
\bibitem{Degrassi:2000qf}
G.~Degrassi, P.~Gambino and G.~F.~Giudice,
%``B $\to$ X/s gamma in supersymmetry: Large contributions beyond the
%leading order,''
JHEP {\bf 0012}, 009 (2000)
[hep-ph/0009337].
%%CITATION = HEP-PH 0009337;%%

\bibitem{Carena:2000uj}
A.~Ali, G.~F.~Giudice and T.~Mannel,
%``Towards a model independent analysis of rare B decays,''
hep-ph/9410289.
%%CITATION = HEP-PH 9410289;%%
M.~Carena, D.~Garcia, U.~Nierste and C.~E.~Wagner,
%``b $\to$ s gamma and supersymmetry with large tan(beta),''
Phys.\ Lett.\ B {\bf 499}, 141 (2001)
[hep-ph/0010003].
%%CITATION = HEP-PH 0010003;%%
K.~i.~Okumura and L.~Roszkowski,
%``$B \to X_s \gamma$ in minimal supersymmetric standard model with general 
%flavor mixing,''
[hep-ph/0212007].
%%CITATION = HEP-PH 0212007;%%
D.~A.~Demir and K.~A.~Olive,
%``B $\to$ X/s gamma in supersymmetry with explicit CP violation,''
Phys.\ Rev.\ D {\bf 65}, 034007 (2002)
[hep-ph/0107329].
%%CITATION = HEP-PH 0107329;%%




\bibitem{Cho:1996we}
P.~L. Cho, M.~Misiak, and D.~Wyler,
\newblock Phys. Rev. {\bf D54}, 3329 (1996), hep-ph/9601360.
%%CITATION = HEP-PH 9601360;%%

\bibitem{Hewett:1997ct}
J.~L. Hewett and J.~D. Wells,
\newblock Phys. Rev. {\bf D55}, 5549 (1997), hep-ph/9610323.
%%CITATION = HEP-PH 9610323;%%



\bibitem{Ali:2002jg}
A.~Ali, E.~Lunghi, C.~Greub and G.~Hiller,
%``Improved model-independent analysis of semileptonic and radiative rare  B 
%decays,''
Phys.\ Rev.\ D {\bf 66}, 034002 (2002)
[hep-ph/0112300].
%%CITATION = HEP-PH 0112300;%%

\bibitem{susybsll3}
E.~Lunghi, A.~Masiero, I.~Scimemi and L.~Silvestrini,
%``B $\to$ X/s l+ l- decays in supersymmetry,''
Nucl.\ Phys.\ B {\bf 568}, 120 (2000)
[hep-ph/9906286].
%%CITATION = HEP-PH 9906286;%%
E.~Lunghi and I.~Scimemi,
%``CP violation in rare semileptonic B decays and supersymmetry,''
Nucl.\ Phys.\ B {\bf 574}, 43 (2000)
[hep-ph/9912430].
%%CITATION = HEP-PH 9912430;%%
C.~S.~Huang and W.~Liao,
%``The electric dipole moment and CP violation B $\to$ X/s l+ l- in  SUGRA 
%models with nonuniversal gaugino masses,''
Phys.\ Rev.\ D {\bf 62}, 016008 (2000).
%%CITATION = HEP-PH 0001174;%%
Z.~Xiong and J.~M.~Yang,
%``Rare B-Meson Dileptonic Decays In Minimal Supersymmetric Model,''
Nucl.\ Phys.\ B {\bf 628}, 193 (2002)
[hep-ph/0105260].
%%CITATION = HEP-PH 0105260;%%
E.~Gabrielli and S.~Khalil,
%``On the B $\to$ X/s l+ l- decays in general supersymmetric models,''
Phys.\ Lett.\ B {\bf 530}, 133 (2002)
[hep-ph/0201049].
%%CITATION = HEP-PH 0201049;%%



\bibitem{expbsgamma}
R.~Barate {\it et al.}  [ALEPH Collaboration],
%``A measurement of the inclusive b $\to$ s gamma branching ratio,''
Phys.\ Lett.\ B {\bf 429} (1998) 169;
%%CITATION = PHLTA,B429,169;%%
%
K.~Abe {\it et al.}  [Belle Collaboration],
%``A measurement of the branching fraction for the inclusive B $\to$ X/s  gamma 
Phys.\ Lett.\ B {\bf 511} (2001) 151
[hep-ex/0103042];
%%CITATION = HEP-EX 0103042;%%
%
S.~Chen {\it et al.}  [CLEO Collaboration],
%``Branching fraction and photon energy spectrum for b $\to$ s gamma,''
Phys.\ Rev.\ Lett.\  {\bf 87} (2001) 251807
[hep-ex/0108032];
%%CITATION = HEP-EX 0108032;%%
%
B.~Aubert {\it et al.}  [BaBar Collaboration],
%``Determination of the branching fraction for inclusive decays B $\to$ X/s  
%gamma,''
hep-ex/0207074,
%%CITATION = HEP-EX 0207074;%%
hep-ex/0207076.
%%CITATION = HEP-EX 0207076;%%



%%%%%%%%%%%%%%%%%%%%%%%%%%%%%%%%%%%%%%%%%%%%%%%%%%%%%%%%%%%%%%%%%%%%%%%%




\bibitem{Bertolini:1998hp}
G.~L.~Lin, J.~Liu and Y.~P.~Yao,
%``Flavor Changing Two Photon Decay,''
Phys.\ Rev.\ Lett.\  {\bf 64}, 1498 (1990).
%%CITATION = PRLTA,64,1498;%%
C.~H.~Chang, G.~L.~Lin and Y.~P.~Yao,
%``QCD corrections to b $\to$ s gamma gamma and exclusive B/s $\to$ gamma gamma decay,''
Phys.\ Lett.\ B {\bf 415}, 395 (1997)
[hep-ph/9705345].
%%CITATION = HEP-PH 9705345;%%
L.~Reina, G.~Ricciardi and A.~Soni,
%``QCD corrections to b $\to$ s gamma gamma induced decays:  B $\to$ X/s gamma 
%gamma and B/s $\to$ gamma gamma,''
Phys.\ Rev.\ D {\bf 56}, 5805 (1997)
[hep-ph/9706253];
%%CITATION = HEP-PH 9706253;%%
Phys.\ Lett.\ B {\bf 396}, 231 (1997)
[hep-ph/9612387].
%%CITATION = HEP-PH 9612387;%%
S.~Bertolini and J.~Matias,
\newblock Phys. Rev. {\bf D57}, 4197 (1998), hep-ph/9709330.
%%CITATION = HEP-PH 9709330;%%
S.~W.~Bosch and G.~Buchalla,
%``The double radiative decays B $\to$ gamma gamma in the heavy quark limit,''
JHEP {\bf 0208}, 054 (2002)
[hep-ph/0208202].
%%CITATION = HEP-PH 0208202;%%


%\cite{Walsh:2002ux}
\bibitem{Walsh:2002ux}
J.~J.~Walsh  [BABAR Collaboration],
%``BABAR results on the decays B $\to$ K l+ l- and B $\to$ K* l+ l-,''
PRINT-02-18
{\it Contributed to Flavor Physics and CP Violation (FPCP),
  Philadelphia, Pennsylvania, }

%\cite{Aubert:2002pj}
\bibitem{Aubert:2002pj}
B.~Aubert {\it et al.}  [BABAR Collaboration],
%``Evidence for the flavor changing neutral current decays B $\to$ K l+ l-  and B $\to$ K* l+ l-,''
hep-ex/0207082.
%%CITATION = HEP-EX 0207082;%%




\bibitem{BBNS}
%\bibitem{Beneke:1999br}
M.~Beneke, G.~Buchalla, M.~Neubert, and C.~T. Sachrajda,
\newblock Phys. Rev. Lett. {\bf 83} (1999) 1914 [hep-ph/9905312];
%%CITATION = HEP-PH 9905312;%%
%
%\bibitem{Beneke:2000ry}
%M.~Beneke, G.~Buchalla, M.~Neubert, and C.~T. Sachrajda,
\newblock Nucl. Phys. {\bf B591} (2000) 313 [hep-ph/0006124];
%%CITATION = HEP-PH 0006124;%%
%
%\bibitem{Beneke:2001ev}
%M.~Beneke, G.~Buchalla, M.~Neubert, and C.~T. Sachrajda,
\newblock Nucl. Phys. {\bf B606} (2001) 245 [hep-ph/0104110].
%%CITATION = HEP-PH 0104110;%%


\bibitem{Beneke:2001at}
M.~Beneke, T.~Feldmann, and D.~Seidel,
\newblock Nucl. Phys. {\bf B612} (2001) 25 [hep-ph/0106067].
%%CITATION = HEP-PH 0106067;%%

\bibitem{Burdman:1998mk}
G.~Burdman,
\newblock Phys. Rev. {\bf D57} (1998) 4254 [hep-ph/9710550].
%%CITATION = HEP-PH 9710550;%%

\bibitem{Ali:1999mm}
A.~Ali, P.~Ball, L.~T. Handoko, and G.~Hiller,
\newblock Phys. Rev. {\bf D61} (2000) 074024 [hep-ph/9910221].
%%CITATION = HEP-PH 9910221;%%


\bibitem{AFB}
W.~Jaus and D.~Wyler,
%``The Rare Decays Of B $\to$ K Lepton Anti-Lepton And B $\to$ K* Lepton Anti-Lepton,''
Phys.\ Rev.\ D {\bf 41}, 3405 (1990).
%%CITATION = PHRVA,D41,3405;%%
P.~Colangelo, F.~De Fazio, P.~Santorelli and E.~Scrimieri,
%``QCD Sum Rule Analysis of the Decays $B \to K \ell~+ \ell~-$ and $B \to K~* \ell~+ \ell~-$,''
Phys.\ Rev.\ D {\bf 53}, 3672 (1996)
[Erratum-ibid.\ D {\bf 57}, 3186 (1998)]
[hep-ph/9510403].
%%CITATION = HEP-PH 9510403;%%
T.~M.~Aliev, A.~Ozpineci and M.~Savci,
%``Rare B $\to$ K* l+ l- decay in light cone QCD,''
Phys.\ Rev.\ D {\bf 56}, 4260 (1997)
[hep-ph/9612480].
%%CITATION = HEP-PH 9612480;%%
D.~Melikhov, N.~Nikitin and S.~Simula,
%``Lepton asymmetries in exclusive b $\to$ s l+ l- decays as a test of the  standard model,''
Phys.\ Lett.\ B {\bf 430}, 332 (1998)
[hep-ph/9803343].
%%CITATION = HEP-PH 9803343;%%

\bibitem{Kagan:2001zk}
A.~L. Kagan and M.~Neubert,
%``Isospin breaking in B $\to$ K* gamma decays,''
Phys.\ Lett.\ B {\bf 539} (2002) 227
[hep-ph/0110078].
%%CITATION = HEP-PH 0110078;%%




\bibitem{Beneke:2000wa}
M.~Beneke and T.~Feldmann,
\newblock Nucl. Phys. {\bf B592} (2001) 3 [hep-ph/0008255].
%%CITATION = HEP-PH 0008255;%%

\bibitem{Bosch:2001gv}
S.~W. Bosch and G.~Buchalla,
\newblock Nucl. Phys. {\bf B621} (2002) 459 [hep-ph/0106081];
%%CITATION = HEP-PH 0106081;%%
%\cite{Bosch:2002bw}
%\bibitem{Bosch:2002bw}
S.~W.~Bosch,
%``Exclusive radiative decays of B mesons in QCD factorization,''
hep-ph/0208203.
%%CITATION = HEP-PH 0208203;%%

\bibitem{Ali:2001ez}
A.~Ali and A.~Y. Parkhomenko,
%``Branching ratios for B $\to$ rho gamma decays in next-to-leading order in  alpha(s) including hard spectator corrections,''
Eur.\ Phys.\ J.\ C {\bf 23} (2002) 89
[hep-ph/0105302].
%%CITATION = HEP-PH 0105302;%%


%\cite{Bosch:2002bv}
\bibitem{Bosch:2002bv}
S.~W.~Bosch and G.~Buchalla,
%``The double radiative decays B $\to$ gamma gamma in the heavy quark limit,''
JHEP {\bf 0208} (2002) 054
[hep-ph/0208202].
%%CITATION = HEP-PH 0208202;%%


%\cite{Descotes-Genon:2002mw}
\bibitem{Descotes-Genon:2002mw}
S.~Descotes-Genon and C.~T.~Sachrajda,
%``Factorization, the light-cone distribution amplitude of the B-meson and  the radiative decay B $\to$ gamma l nu/l,''
hep-ph/0209216.
%%CITATION = HEP-PH 0209216;%%



%\cite{Ali:2002qc}
\bibitem{Ali:2002qc}
A.~Ali and A.~S.~Safir,
%``Helicity analysis of the decays B $\to$ K* l+ l- and B $\to$ rho l nu/l  in the large energy effective theory,''
hep-ph/0205254.
%%CITATION = HEP-PH 0205254;%%



\bibitem{BEneke:1998rk}
M.~Beneke,
\newblock Phys. Lett. {\bf B434} (1998) 115 [hep-ph/9804241].
%%CITATION = HEP-PH 9804241;%%


\bibitem{Charles:1998dr}
J.~Charles, A.~Le~Yaouanc, L.~Oliver, O.~P\`ene, and J.~C. Raynal,
\newblock Phys. Rev. {\bf D60} (1999) 014001 [hep-ph/9812358].
%%CITATION = HEP-PH 9812358;%%


\bibitem{Descotes-Genon:2001hm}
S.~Descotes-Genon and C.~T. Sachrajda,
\newblock Nucl. Phys. {\bf B625} (2002) 239 [hep-ph/0109260].
%%CITATION = HEP-PH 0109260;%%



\bibitem{Grozin:1997pq}
A.~G. Grozin and M.~Neubert,
\newblock Phys. Rev. {\bf D55} (1997) 272 [hep-ph/9607366].
%%CITATION = HEP-PH 9607366;%%


\bibitem{Korchemsky:1999qb}
G.~P. Korchemsky, D.~Pirjol, and T.-M. Yan,
\newblock Phys. Rev. {\bf D61} (2000) 114510 [hep-ph/9911427].
%%CITATION = HEP-PH 9911427;%%

\bibitem{Khodjamirian:2001ga}
A.~Khodjamirian and D.~Wyler,
\newblock (2001), hep-ph/0111249.
%%CITATION = HEP-PH 0111249;%%

\bibitem{Beyer:2001zn}
M.~Beyer, D.~Melikhov, N.~Nikitin, and B.~Stech,
\newblock Phys. Rev. {\bf D64} (2001) 094006 [hep-ph/0106203].
%%CITATION = HEP-PH 0106203;%%

\bibitem{Dincer:2001hu}
Y.~Dincer and L.~M.~Sehgal,
%``Charge asymmetry and photon energy spectrum in the decay B/s $\to$  l+ l- gamma,''
Phys.\ Lett.\ B {\bf 521}, 7 (2001)
[hep-ph/0108144].
%%CITATION = HEP-PH 0108144;%%



\bibitem{Grinstein:2000pc}
B.~Grinstein and D.~Pirjol,
\newblock Phys. Rev. {\bf D62} (2000) 093002 [hep-ph/0002216].
%%CITATION = HEP-PH 0002216;%%

%\cite{Kruger:2002gf}
\bibitem{Kruger:2002gf}
F.~Kr{\"u}ger and D.~Melikhov,
%``Gauge invariance and form factors for the decay B $\to$ gamma l+ l-,''
hep-ph/0208256.
%%CITATION = HEP-PH 0208256;%%

%\cite{Lunghi:2002ju}
\bibitem{Lunghi:2002ju}
E.~Lunghi, D.~Pirjol and D.~Wyler,
%``Factorization in leptonic radiative B $\to$ gamma e\nu decays,''
hep-ph/0210091.
%%CITATION = HEP-PH 0210091;%%

\bibitem{Melikhov:2001sd}
D.~Melikhov and B.~Stech,
%``Non-local anomaly of the axial-vector current for bound states,''
Phys.\ Rev.\ Lett.\  {\bf 88} (2002) 151601
[hep-ph/0108165].
%%CITATION = HEP-PH 0108165;%%

\bibitem{Ali:1995uy}
A.~Ali and V.~M. Braun,
\newblock Phys. Lett. {\bf B359} (1995) 223 [hep-ph/9506248].
%%CITATION = HEP-PH 9506248;%%

%\cite{Ball:2002ps}
\bibitem{Ball:2002ps}
P.~Ball, V.~M.~Braun and N.~Kivel,
%``Photon distribution amplitudes in QCD,''
hep-ph/0207307.
%%CITATION = HEP-PH 0207307;%%



\bibitem{Coan:1999kh}
CLEO, T.~E. Coan {\em et~al.},
\newblock Phys. Rev. Lett. {\bf 84}, 5283 (2000), hep-ex/9912057.
%%CITATION = HEP-EX 9912057;%%

\bibitem{Ushiroda:2001sb}
Belle, Y.~Ushiroda,
\newblock (2001), hep-ex/0104045.
%%CITATION = HEP-EX 0104045;%%

\bibitem{Nash:2002vr}
BABAR, J.~Nash,
\newblock (2002), hep-ex/0201002.
%%CITATION = HEP-EX 0201002;%%


\bibitem{list}
A.~Ali and D.~London,
%``Profiles of the unitarity triangle and CP-violating phases in the  standard 
%model and supersymmetric theories,''
Eur.\ Phys.\ J.\ C {\bf 9}, 687 (1999)
[hep-ph/9903535].
%%CITATION = HEP-PH 9903535;%%
A.~J.~Buras, P.~Gambino, M.~Gorbahn, S.~J{\"a}ger and 
L.~Silvestrini,
%``Universal unitarity triangle and physics beyond the standard model,''
Phys.\ Lett.\ B {\bf 500}, 161 (2001)
[hep-ph/0007085].
%%CITATION = HEP-PH 0007085;%%
A.~J.~Buras and R.~Fleischer,
%``Bounds on the unitarity triangle, sin(2beta) and K $\to$ pi nu anti-nu  
%decays in models with minimal flavor violation,''
Phys.\ Rev.\ D {\bf 64}, 115010 (2001)
[hep-ph/0104238].
%%CITATION = HEP-PH 0104238;%%
A.~Ali and E.~Lunghi,
%``Extended minimal flavor violating MSSM and implications for B physics,''
Eur.\ Phys.\ J.\ C {\bf 21}, 683 (2001)
[hep-ph/0105200].
%%CITATION = HEP-PH 0105200;%%
A.~J.~Buras, P.~H.~Chankowski, J.~Rosiek and L.~Slawianowska,
%``Delta(M(s))/Delta(M(d)), sin 2beta and the angle gamma in the presence  of 
%new Delta(F) = 2 operators,''
Nucl.\ Phys.\ B {\bf 619}, 434 (2001)
[hep-ph/0107048].
%%CITATION = HEP-PH 0107048;%
S.~Laplace, Z.~Ligeti, Y.~Nir and G.~Perez,
%``Implications of the CP asymmetry in semileptonic B decay,''
Phys.\ Rev.\ D {\bf 65}, 094040 (2002)
[hep-ph/0202010].
%%CITATION = HEP-PH 0202010;%%

\bibitem{amb}
G.~D'Ambrosio, G.~F.~Giudice, G.~Isidori and A.~Strumia,
%``Minimal flavour violation: An effective field theory approach,''
hep-ph/0207036.
%%CITATION = HEP-PH 0207036;%%

\bibitem{private}
G.~Degrassi and P.~Gambino, private communication.


\bibitem{Altarelli:1997et}
G.~Altarelli, R.~Barbieri and F.~Caravaglios,
%``Electroweak precision tests: A concise review,''
Int.\ J.\ Mod.\ Phys.\ A {\bf 13}, 1031 (1998)
[hep-ph/9712368];
%%CITATION = HEP-PH 9712368;%%
J.~Erler and D.~M.~Pierce,
%``Bounds on supersymmetry from electroweak precision analysis,''
Nucl.\ Phys.\ B {\bf 526}, 53 (1998)
[hep-ph/9801238];
%%CITATION = HEP-PH 9801238;%%
A.~Djouadi, P.~Gambino, S.~Heinemeyer, W.~Hollik, C.~Junger and 
G.~Weiglein,
%``Leading {QCD} corrections to scalar quark contributions to electroweak precision observables,''
Phys.\ Rev.\ D {\bf 57}, 4179 (1998)
[hep-ph/9710438];
%%CITATION = HEP-PH 9710438;%%
P.~H.~Chankowski and S.~Pokorski,
%``Chargino Mass and $R_b$ Anomaly,''
Nucl.\ Phys.\ B {\bf 475}, 3 (1996)
[hep-ph/9603310].
%%CITATION = HEP-PH 9603310;%%
G.~Altarelli, F.~Caravaglios, G.~F.~Giudice, P.~Gambino and G.~Ridolfi,
%``Indication for light sneutrinos and gauginos from precision electroweak  data,''
JHEP {\bf 0106}, 018 (2001)
[hep-ph/0106029].
%%CITATION = HEP-PH 0106029;%%



%\cite{Safir:2002gc}
\bibitem{Safir:2002gc}
A.~S.~Safir,
%``Theoretical studies of exclusive rare B decays in the Standard Model
% and Supersymmetric theories,''
hep-ph/0211103.
%%CITATION = HEP-PH 0211103;%%



%\cite{Buchalla:2000sk}
\bibitem{Buchalla:2000sk}
G.~Buchalla, G.~Hiller and G.~Isidori,
%``Phenomenology of non-standard Z couplings in exclusive semileptonic
%  b $\to$ s transitions,''
Phys.\ Rev.\ D {\bf 63}, 014015 (2001)
[hep-ph/0006136].
%%CITATION = HEP-PH 0006136;%%

%\cite{Grossman:1999av}
\bibitem{Grossman:1999av}
Y.~Grossman, M.~Neubert and A.~L.~Kagan,
%``Trojan penguins and isospin violation in hadronic B decays,''
JHEP {\bf 9910}, 029 (1999)
[hep-ph/9909297].
%%CITATION = HEP-PH 9909297;%%

\end{thebibliography}
\end{document}